\documentclass{article}

\usepackage{arxiv}

\usepackage[utf8]{inputenc} 
\usepackage[T1]{fontenc}    
\usepackage{hyperref}       
\usepackage{url}            
\usepackage{booktabs}       
\usepackage{amsfonts}       
\usepackage{nicefrac}       
\usepackage{microtype}      
\usepackage{lipsum}		
\usepackage{graphicx}
\usepackage{doi}

 \usepackage{multirow}

\usepackage{ mathrsfs }

\usepackage{lineno,hyperref}
\usepackage{xcolor}

\usepackage{amsfonts}

\usepackage{graphicx}
\usepackage{subfig}
\usepackage{pdflscape}
\usepackage[fleqn]{amsmath}
\usepackage{breqn}
\usepackage{bigints}
\usepackage{epstopdf}
\usepackage{tikz}
\usetikzlibrary{matrix,calc}

\usepackage{accents}

\usepackage{comment}

\usepackage{algorithm}
\usepackage{algpseudocode}

\usepackage{hyperref}
\hypersetup{
    colorlinks=true,
    citecolor=red,
    linkcolor=blue,
    filecolor=magenta,      
    urlcolor=cyan,
}

\usepackage{titlesec}

\titleformat{\paragraph}
{\normalfont\normalsize\bfseries}{\theparagraph}{1em}{}
\titlespacing*{\paragraph}
{0pt}{3.25ex plus 1ex minus .2ex}{1.5ex plus .2ex}

\usepackage{xcolor}
\usepackage{soul}

\usepackage{booktabs,caption}
\usepackage[flushleft]{threeparttable}

\title{Edge Ranking of Graphs in Transportation Networks using a Graph Neural Network (GNN)}

\author{ \href{https://orcid.org/0000-0003-2368-6394}{\includegraphics[scale=0.06]{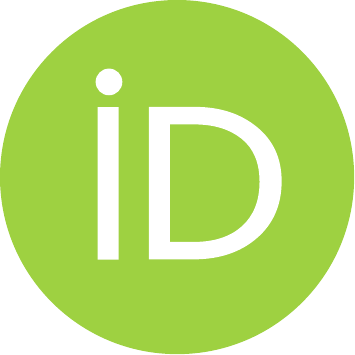}\hspace{1mm}Debasish Jana}\thanks{Authors' with equal contribution} \\
	Samueli Civil and Environmental Engineering\\
	University of California Los Angeles\\
	Los Angeles, CA-90095 \\
	\texttt{dj93@ucla.edu} \\
	\And
	\href{https://orcid.org/0000-0003-2552-1147}{\includegraphics[scale=0.06]{orcid.pdf}\hspace{1mm}Sven Malama}\footnotemark[1] \\
	Samueli Civil and Environmental Engineering\\
	University of California Los Angeles\\
	Los Angeles, CA-90095 \\
	\texttt{svenmala@g.ucla.edu} \\
	\AND
	\href{https://orcid.org/0000-0003-0412-6244}{\includegraphics[scale=0.06]{orcid.pdf}\hspace{1mm}Sriram Narasimhan}\thanks{Corresponding Author} \\ 
	Samueli Civil and Environmental Engineering \\
    Samueli Mechanical and Aerospace Engineering \\
    University of California Los Angeles \\
	Los Angeles, CA-90095 \\
	\texttt{snarasim@g.ucla.edu} \\
	\And
	\href{https://orcid.org/0000-0001-9618-1210}{\includegraphics[scale=0.06]{orcid.pdf}\hspace{1mm}Ertugrul Taciroglu} \\
	Samueli Civil and Environmental Engineering \\
    University of California Los Angeles \\
	Los Angeles, CA-90095 \\
	\texttt{etacir@g.ucla.edu} \\
}

\begin{document}
\maketitle

\begin{abstract}
Many networks, such as transportation, power, and water distribution, can be represented as graphs. A crucial challenge in graph representations is identifying the importance of graph edges and their influence on the overall performance in terms of network efficiency and information flow. For example, important edges in a transportation network are those roads that, when affected, will significantly alter the network's overall efficiency. A commonly used approach to finding such important edges is ``edge betweenness centrality'' (EBC)---an edge ranking measure to determine the influential edges of the graph based on connectivity and information spread. Computing the EBC utilizing the common Brandes algorithm \cite{brandes2001faster} involves calculating the shortest paths for \emph{every} node pair, which can be computationally expensive and restrictive, especially for large graphs. Changes in the graph parameters, e.g., in the edge weight or the addition and deletion of nodes or edges, require the recalculation of the EBC. As the main contribution, we propose an approximate method to estimate the EBC using a Graph Neural Network (GNN), a deep learning-based approach. We show that it is computationally efficient compared to the conventional method, especially for large graphs. The proposed method of GNN-based edge ranking is evaluated on several synthetic graphs and a real-world transportation data set. We show that this framework can estimate the approximate edge ranking much faster compared to the conventional method introduced by Brandes \cite{brandes2001faster}. This approach is inductive---i.e., training and testing are performed on different sets of graphs with varying numbers of nodes and edges. The proposed method is especially suitable for applications on large-scale networks when edge information is desired, for example, in urban infrastructure improvement projects, power and water network resilience analyses, and optimizing resource allocations in engineering networks.
\end{abstract}


\section{Introduction}
\subsection{Motivation}

\noindent
Transportation networks play a crucial role in the economy and the well-being of citizens by enabling the smooth movement of people and goods and as arteries for evacuations during catastrophes and natural disasters. A healthy transportation network offers significant benefits to its citizens, e.g., through the mobility of capital and labor, diffusion of population, or national defense \cite{Important2022}. Natural hazard events can severely impact transportation networks leading to direct losses, such as repair costs of infrastructure, and indirect losses, such as a decrease in network efficiency \cite{Postance2017Extending}. In 2005, Hurricane Katrina severely impacted the U.S. highway system, especially in the area along and to the south of the I-10/I-12 corridor. Whereas some elements of the highway system were repaired and re-initiated in weeks, other elements remained impassible for many months \cite{Grenzeback2008CaseSO}. Even two years after the disaster, basic services in New Orleans and public transportation and libraries did not regain half of its pre-Katrina capacity \cite{Liu2007}.

A good understanding of a transportation network's performance, capacity, and critical road segments are essential in decision-making during such catastrophic events. One of the critical tasks related to such an understanding is to objectively identify which road segments are crucial to the system's performance as a whole. A powerful means to answer this is to treat the network as a graph comprised of nodes and edges, representing road junctions and segments sections respectively, and optimizing for key measures such as travel-time cost or distance using this graph by associating weights to edges. Such nodes or edges can be deleted, or edge weights modified to simulate loss or decrease of functionality in local regions of the network. 
Node ranking provides a natural application in social networks \cite{ZAREIE2019120}; on the other hand, edge importance ranking is more suited for engineered network systems, such as transportation networks. Identifying critical edges in transportation networks can affect preemptive strategies to address deficiencies in essential segments of the system, thereby making the overall system more robust and resilient to failures. The central objective of this paper is to propose a novel and computationally efficient way to estimate the importance of edges in transportation networks. This approximate method based on the Graph Neural Network (GNN) is shown to outperform conventional methods in terms of speed while also achieving a comparable level of performance. 


\subsection{Literature Review}
\noindent
Much of the current literature focuses on finding important nodes rather than edges in a graph. Node ranking is relevant in applications such as identifying vulnerable populations for infectious disease spread \cite{brockmann2013hidden, wang2016statistical}, or in social networks \cite{pei2014searching}. For urban infrastructures, edge ranking can be very important, say street sections represented as  edges in a transportation network. A relatively large amount of literature on graph components (nodes and edges) ranking can be found in post-disaster recovery research, where optimal sequencing for the repair of components is necessary to maximize the efficiency/resilience of the network. Vugrin et al. \cite{vugrin2014optimal} presented a bi-level optimization algorithm for optimal recovery sequencing in transportation networks. Gokalp et al. \cite{gokalp2021post} proposed a  bidirectional search heuristic strategy for post-disaster recovery sequencing of bridges for road networks. Bocchini and Frangopol \cite{bocchini2012restoration} presented a model for recovery planning for networks of highway bridges damaged in an earthquake. This model identifies bridge restoration activities that maximizes resilience and minimizes the time required to return the network to a targeted functionality level and the cost of restoration. Network recovery has been studied for other networks such as electrical power restoration \cite{xu2007optimizing}, airline system recovery \cite{clausen2010disruption}, post-earthquake water distribution network recuperation \cite{luna2011postearthquake}, internet protocol (IP) networks rehabilitation \cite{wang2011progressive}. These studies generally cover small search sets of edges in a graph for optimization purposes, assuming only a few roads/bridges are damaged (or, modified in a graph) after a disaster, which is a reasonable assumption. Dealing with a complete network can be computationally exhaustive but might be necessary in case of large scale events like hurricane Katrina, which impacted  15,947 lane miles of highway in Alabama, 60,727 in Louisiana and 28,889 in Mississippi \cite{Hicks2005HurricaneKP}.

The centrality measure is a metric used for ranking nodes or edges \cite{dehmer2014quantitative}. This metric represents a quantitative view of how a component's absence or presence affects the whole graph. The well-studied importance metrics for nodes are: the betweenness centrality \cite{freeman1977set, newman2005measure}, closeness centrality \cite{bavelas1950communication, sabidussi1966centrality}, and page-rank centrality \cite{brin1998anatomy, ding2009pagerank, franceschet2011pagerank, gleich2015pagerank}. The aforementioned centrality measures are mainly designed for node ranking. For example, betweenness centrality is a measure of the amount of network information flow being controlled by a specific node \cite{borgatti1995centrality}. Brohl et al. \cite{brohl2019centrality} modified the formula so that centralities can be computed for edges. The most commonly used metric for importance estimation of edge-components is edge betweenness centrality (EBC) \cite{brandes2001faster, brandes2008variants,freeman1977set, girvan2002community}. EBC is based on how the edges expedite the flow of information in a graph. Edges with higher EBCs are considered high-importance edges, where the removal of an edge with large EBC significantly disrupts the information flow in a graph. \cite{borgatti1995centrality}.

The application of node betweenness centrality can be found in the study of biological graphs \cite{joy2005high}, contingency analysis in power grids \cite{jin2010novel}, knowledge network analysis \cite{pham2010structure}, and traffic monitoring in transportation networks \cite{puzis2013augmented}. Betweenness centrality calculation for both nodes and edges requires the shortest path estimation from each node to every other node in the graph. Therefore, the calculation of the betweenness centrality score is computationally expensive, especially for large graphs. An approximate calculation of the centrality score based on sampling techniques can overcome this challenge. Geisberger et al. \cite{geisberger2008better} proposed a method to estimate the approximate betweenness centrality of $k$ nodes sampled randomly to find the shortest path and calculate the betweenness centrality of all nodes. Riondato et al. \cite{riondato2014fast} chose $k$ shortest paths between randomly sampled source-target node pairs and evaluated the betweenness centrality for all nodes. Borassi \cite{borassi2019kadabra} proposed adaptive sampling techniques for sampling shortest paths to compute betweenness centralities faster. Mahmoody et al. \cite{mahmoody2016scalable} studied centrality maximization problems for graphs and proposed an efficient randomized algorithm to approximate the node centrality score for larger graphs. Yoshida \cite{yoshida2014almost} proposed a hypergraph-based approach to estimate adaptive betweenness centrality for dynamic graphs. However, these random-sampling-based approximate algorithms lead to sub-optimal ranking accuracy and an increase in execution time for extensive and dynamically changing networks \cite{newman2006modularity}. 

Recent advancements in computing and the availability of large data sets have resulted in powerful machine learning and deep learning methods. Mendon{\c{c}}a et al. \cite{mendoncca2020approximating} proposed a simple neural network with graph embedding to estimate the approximate node betweenness centrality. A Graph Neural Network (GNN) is a deep learning architecture that leverages the graph structure and feature information to perform various tasks including node/edge/graph classification \cite{zhou2020graph}. For transportation networks, GNNs have been used in traffic demand prediction \cite{liang2022joint} and traffic speed forecasting \cite{yu2020forecasting}.  Maurya et al. \cite{maurya2019fast, maurya2021graph} proposed a GNN-based node ranking framework, and Fan et al. \cite{fan2019learning} used GNNs to determine high-importance nodes. Park et al. \cite{park2019estimating} adopted GNNs to estimate node importance in knowledge graphs. The aforementioned GNN-based methods exclusively work on node centrality ranking on unweighted directed and undirected graphs.

\subsection{Contributions}
\noindent
Current literature primarily focuses on approximate node betweenness centrality for unweighted graphs using GNNs and other sampling methods; approximate estimation EBC for large graphs is lacking. Edge importance ranking is essential in dealing with problems relevant to urban infrastructure networks, such as transportation or utility distribution networks. The primary contribution of this paper is a fast and accurate GNN-based approximate edge ranking approach for weighted graphs. Changes in edge weights, restructuring of nodes and edge formation, and node/edge failures can lead to differences in edge importance rankings. Recalculating these edge importance rankings using the conventional method is time-consuming, especially for large-scale graphs. The proposed GNN-based approach reduces the computational time significantly by exploiting both the inherent parallelism of neural networks and GPUs for large-scale matrix operations along the same lines as deep learning techniques \cite{lecun2015deep, goodfellow2016deep}. The main principle of GNNs is to aggregate node features of the neighboring connected nodes in the graph. In multi-layer GNNs, such repetitive aggregation captures the overall structural and neighborhood information of the node of interest. The proposed method modifies the conventional GNN architecture to work on edges instead of nodes. Specifically, the proposed method uses a modified edge-adjacency matrix of the graph that is assimilated using the node-degree and edge-weight information to estimate the edge ranking accurately. This modification to the original edge-adjacency matrix leads to a unique representation. To the authors' knowledge, this is the first work where GNN is used to approximate the EBC in transportation networks. The trained GNN model can perform edge ranking approximation for static and dynamic graphs (time-varying graph systems). Its performance is demonstrated on synthetically generated graphs and a real-world transportation network.

\subsection{Organization}

\noindent
The remainder of this paper is organized as follows. First, in Section \ref{Sec_Prelim}, we present the basics of graph theory, EBC, and the computation of edge feature representation. Next, the working principles of GNN and information propagation in the learning stage are described. Subsequently, in Section \ref{Sec_proposedFramework}, we present the new GNN-based framework, which forms the core of the contributions claimed in this paper. In Section \ref{Sec_Results}, we evaluate the performance of the proposed approach on synthetic graphs. Next, we demonstrate the performance on a real-world transportation network in Section \ref{Sec_Results}. Finally, the conclusions are presented in Section \ref{Sec_Conclusion}. 


\section{Preliminaries}\label{Sec_Prelim}

This section introduces the background on the concepts and terminologies necessary to follow the material presented in this paper. The basics of graph theory are explained along with an introduction to the edge adjacency matrix, which describes the spatial connection between edges in a graph. Then, a brief introduction to the conventional method of computing EBC, including its drawbacks, follows. Next, we briefly introduce edge feature representation in the graph topology. Finally, the basic concepts of Graph Neural Networks (GNNs), including how the information of edges is exchanged and accumulated, are described. The original GNN algorithm \cite{wu2020comprehensive} proposes the message passing on nodes, whereas here, the message passing is on edges. Through the edge adjacency matrix and the edge feature vectors, the GNN learns to predict an approximate edge rank.

\subsection{Basics of Graph theory}

A graph $\mathcal{G}, $ is defined as $(\mathcal{V}, \mathcal{E})$ -- here $\mathcal{V}$ denotes the set of nodes or vertices of the graph and $\mathcal{E} \subseteq \mathcal{V} \times \mathcal{V}$ symbolizes the edges \cite{godsil2001algebraic}. The neighbor set of node $i \in \mathcal{V}$ is defined as $\mathcal{N}_i := \{ j \in \mathcal{V} : (i,j) \in \mathcal{E} \}$. The graph edges are weighted by $w_{ij}$ which are associated with $(i,j)$ for $i,j \in \mathcal{V}$ -- here $w_{ij} > 0$ if $(i,j) \in \mathcal{E}$ and $w_{ij} = 0$ otherwise. The vertex adjacency matrix (or commonly known as adjacency matrix) $\mathcal{A}^{\mathcal{V}} = [a^{v}_{ij}] \in \mathbb{R}^{|\mathcal{V}| \times |\mathcal{V}|}$ of $\mathcal{G} = (\mathcal{V}, \mathcal{E})$ is defined as \cite{harary1971graph}:

\begin{equation}\label{Eq_NodeAdjMat}
  a^v_{ij} =
    \begin{cases}
      \text{$0$,} & \text{if $i = j$ or there is NO edge present between $i$ and $j$}\,,\\
      \text{$w_{ij}$,} & \text{if $i \neq j$ and there is one edge present between $i$ and $j$\,.}
    \end{cases}       
\end{equation}

Here, the graph $\mathcal{G}$ is an undirected graph such that $(j,i) \in \mathcal{E}$ iff $w_{ij} = w_{ji} \quad \forall \; (i,j) \in \mathcal{E}$. $|\cdot|$ refers to the cardinality, or the number of elements in the set. For unweighted graphs all the weight values are 1 i.e., $  w_{ij} = 1~\forall i,~j$. In the vertex adjacency matrix, non-zero values sparsely appear when an edge exists between two nodes. The edge adjacency matrix $\mathcal{A}^{\mathcal{E}} = [a^{e}_{mn}] \in \mathbb{R}^{|\mathcal{E}| \times |\mathcal{E}|}$ is determined by the adjacency of edges \cite{trinajstic2018chemical, rouvray1976topological}:

\begin{equation}\label{Eq_EdgeAdjMat}
  a^e_{mn} =
    \begin{cases}
      \text{$1$,} & \text{if edges $m$ and $n$ are adjacent}\,,\\
      \text{$0$,} & \text{otherwise}\,.
    \end{cases}       
\end{equation}

\begin{figure}[h!]
    \begin{center}
        \includegraphics[width=0.8\textwidth]{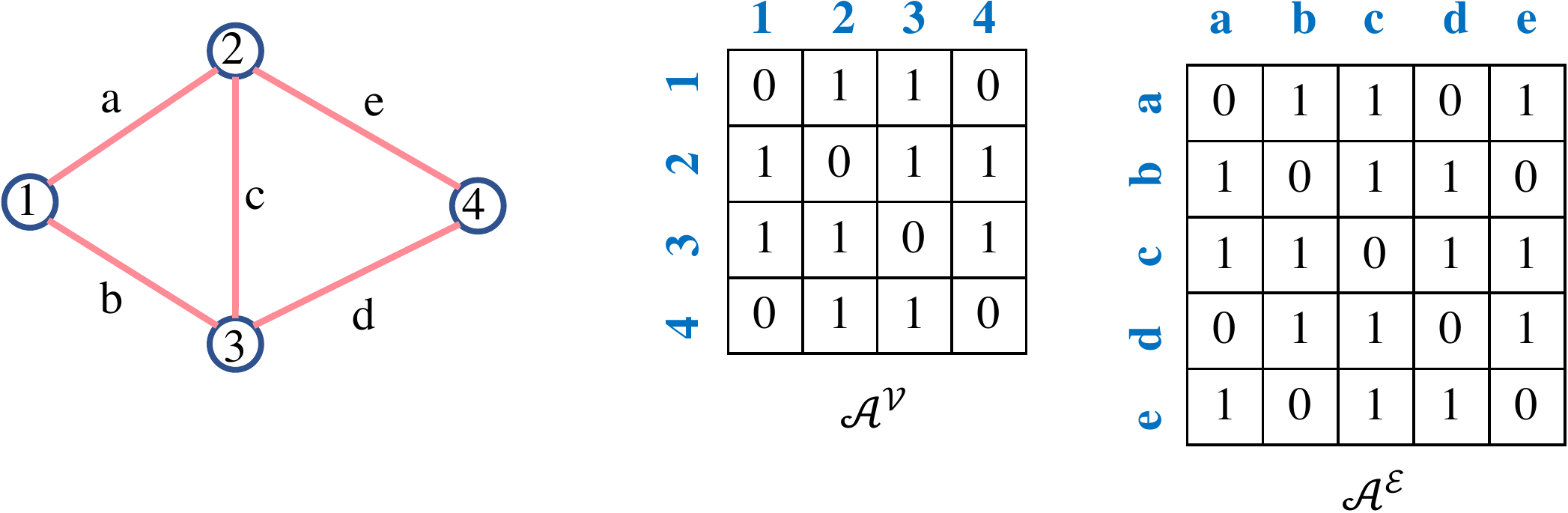}
        \caption{Vertex adjacency matrix and edge adjacency matrix of a sample graph network}
        \label{SampleGraph_Paper1}
    \end{center}
\end{figure}

Figure \ref{SampleGraph_Paper1} shows an example of the vertex adjacency matrix and the edge adjacency matrix for the same graph. 


\subsection{Basics of EBC}\label{Sec_EBC_Conventional}

Edge ranking depends on the edge's ability to control the information flow (the term information flow is contextual) between other nodes and edges of the graph and is highly correlated with the edge weights. The edge weights greatly influence the shortest paths calculated using the graph. Edge ranking based on this criterion is called edge betweenness centrality (EBC) \cite{brandes2001faster, brandes2008variants}. The EBC score of an edge will be high if that edge contains many shortest paths making the information flow more accessible and faster throughout the whole graph. The edges with high betweenness centrality are called `bridges.' Removing bridges from the graph can be disruptive, and in some cases, one graph can segregate into several smaller isolated graphs. Therefore, it is vital to ensure the safety and functionality of such edges in many engineering application contexts (transportation, power, etc.).

For a given graph $\mathcal{G} = (\mathcal{V}, \mathcal{E})$, EBC of an edge $e$ is the sum of the fraction of all-pairs shortest paths that pass through $e$ and is given by \cite{brandes2008variants}:

\begin{equation}\label{Eq_3}
    c_B(e) = \sum_{s,t \in \mathcal{V}} \frac{\sigma(s,t|e)}{\sigma(s,t)}
\end{equation}

\noindent where, $\mathcal{V}$ and $\mathcal{E}$ are the set of nodes and edges, respectively, $ s$ and $t$ are the source and terminal nodes while calculating the shortest paths. $\sigma(s,t)$ is the number of shortest $(s,t)$ paths and $\sigma(s,t|e)$ is the number of those paths passing through edge $e \in \mathcal{E}$ .

The conventional method to calculate this betweenness centrality is through Brandes's algorithm \cite{brandes2001faster}. This algorithm has a space complexity of $\mathcal{O}(|\mathcal{V}|+|\mathcal{E}|)$ and time complexity $\mathcal{O}(|\mathcal{V}||\mathcal{E}|)$ for unweighted networks. For weighted networks, the time complexity increases to $\mathcal{O}(|\mathcal{V}||\mathcal{E}| + |\mathcal{V}|^2 \log{}|\mathcal{V}|)$ \cite{bhardwaj2011performance}. This algorithm is computationally intensive on large-scale networks (examples shown later in Section \ref{Sec_Results}). Additionally, this algorithm is sensitive to small perturbations in the network, such as changes in edge weights or regional node or edge failures. As a result, EBC is recalculated every time there is a change in the graph, which makes its practical implementation in applications such as disaster recovery planning very cumbersome. To address this issue, we pose the estimation of EBC as a learning problem and develop a deep learning-based framework whose time complexity is $\mathcal{O}(|\mathcal{V}|)$ \cite{pimentel2019efficient}.

\subsection{Node and Edge Feature Embeddings}\label{Sec_FeatureRep}
The adjacency matrices represent the connection information between the nodes and edges; however, the complete neighborhood information for nodes and edges is still incomplete beyond their immediate neighbors. The feature representation for nodes and edges embeds the knowledge of $k$-hop neighbors -- hence the information is more exhaustive. Feature representation of the graph components is a way to represent the notion of similarity in graph components. Such embeddings capture the network's topology in a vector format which is crucial for numerical computations and learning. The most popular method for node embedding is \textsf{Node2Vec} \cite{grover2016node2vec}.

\textsf{Node2vec} \cite{grover2016node2vec} is a graph embedding algorithm to transform a graph into a numerical representation. This algorithm generates a feature representation for each node that portrays the whole graph structure, such as node connectivity, weights of the edges, etc. Two similar types of nodes in the graph will have the same numerical representation in \textsf{Node2vec} algorithm. This representation is obtained through second-order biased random walks, and this process is executed in three stages:

\begin{enumerate}
\item {\em First order random walk}

A random walk is a graph traversing procedure along the edges of the graph, best understood by imagining the movement of a walker. First-order random walks sample the nodes on the graph along the graph edges depending on the current state. In each step/hop, the walker transitions from the current state to the next referred to as a 1-hop transition. In Figure \ref{node2vec_ppt2}(a), the walker is at node $v$ and three neighboring nodes are $u_1$, $u_2$, and $u_3$ with the respective edge weights, $w(v,u_1)$, $w(v,u_2)$, and $w(v,u_3)$. These weights determine the probability of the walker transitioning to the next node. The transition probability for the first step is given as,

\begin{equation}
    p(u_i|v)=\frac{w(u_i,v)}{\displaystyle \sum_{u_i \in \mathcal{N}_v} w(u_i,v)} = \frac{w(u_i,v)}{\mathrm{Degree \; of \; node\; }v}; \quad \mathcal{N}_v \; \mathrm{is\; the\; set\; of\; neighboring\; nodes\; of\; } v.
\end{equation}

One random walk is generated by performing multiple one-hop transitions; this process is repeated to multiple random walks, a function of the current state. 

\begin{figure}[h!]
    \begin{center}
        \includegraphics[width=0.9\textwidth]{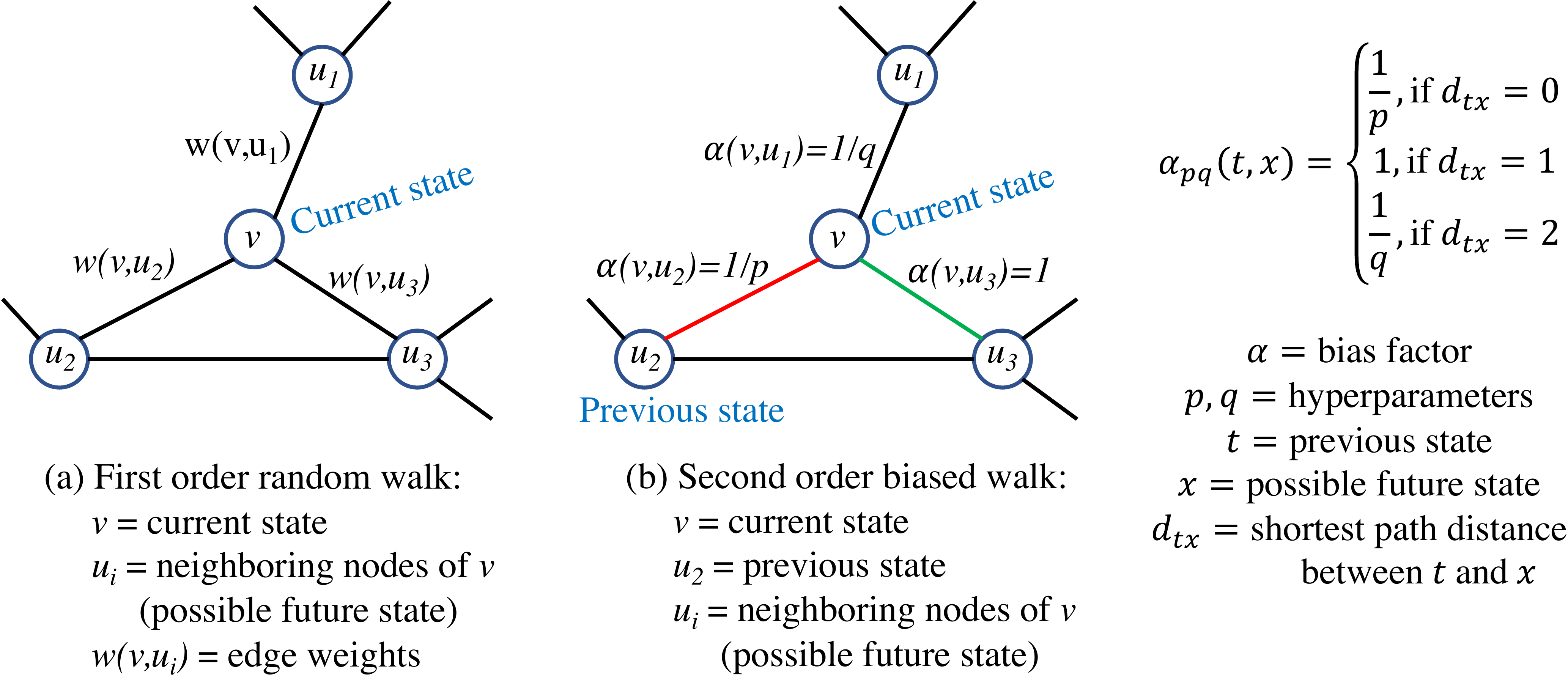}
        \caption{Conceptual representation of \textsf{Node2Vec}: (a) Parameters for transition probability calculation for the 1st order random walk, and (b) parameters for transition probability calculation for the 2nd order biased walk,}
        \label{node2vec_ppt2}
    \end{center}
\end{figure}

\item {\em Second-order biased walk}\label{Sec_SecondOrderWalk}

In the second-order biased walk, the edge weights selection differs from the first-order random walk. A new bias factor term $\alpha$ is introduced to reweigh the edges. The value of $\alpha$ depends on the current state, previous state, and potential future state, as shown in Figure \ref{node2vec_ppt2}(b). If the previous and future states are not connected, then $\alpha = \displaystyle \frac{1}{q}$, $q$ is the in-out parameter. If the previous and future states are identical, then $\alpha = \displaystyle \frac{1}{p}$, where $p$ is the return parameter. If the two states (the previous state and the future state) are connected but not identical, then $\alpha = 1$. Considering the bias factors, the 2\textsuperscript{nd} order transition probability is given as:

\begin{equation}
    p(u_i|v, t)=\frac{\alpha_{pq}(t,u_i) w(u_i,v)}{\displaystyle \sum_{u_i \in \mathcal{N}_v} \alpha_{pq}(t,u_i) w(u_i,v)}
\end{equation}

\newpage
\item {\em Node embeddings from random walks:}

Repeated generation of random walks from every node in the graph results in a large corpus of node sequences. The \textsf{Word2Vec} \cite{mikolov2013distributed} algorithm takes this large corpus as an input to generate the node embeddings. Specifically, \textsf{Node2vec} uses the skip-gram with negative sampling. The main idea of the skip-gram is to maximize the probability of predicting the correct context node given the center node. The skip-gram process for the node embedding is shown in Figure \ref{word2vec_ppt2}.

\begin{figure}[h!]
    \begin{center}
        \includegraphics[width=\textwidth]{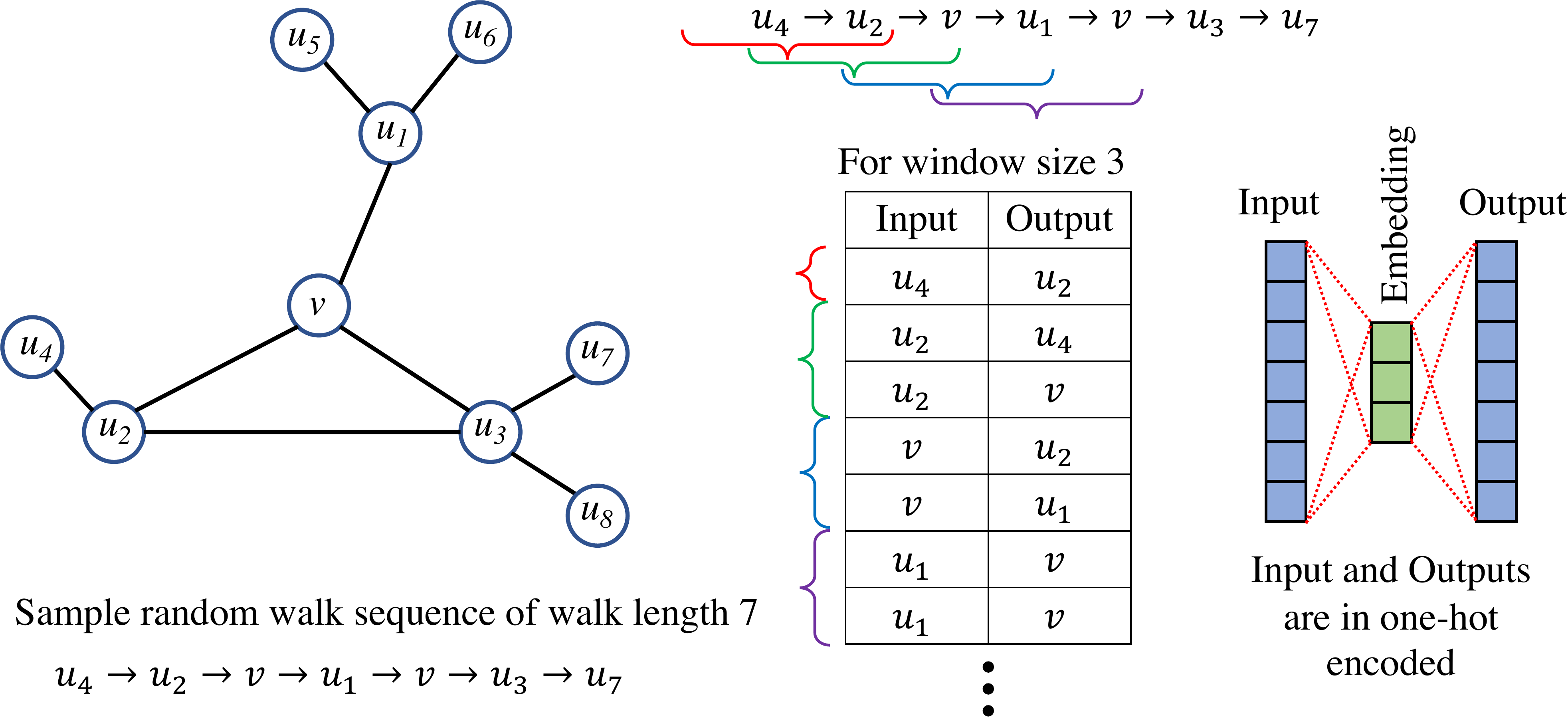}
        \caption{This is an illustration for the skipgram model for node embedding. For a sample random walk of length 7, a sliding window of length 3 is used to prepare the inputs and outputs for training the \textsf{Word2Vec} model. The embedding of the trained \textsf{Word2Vec} model is the node feature embedding.}
        \label{word2vec_ppt2}
    \end{center}
\end{figure}

From the node embedding, the edge embedding is obtained using the average operator---edge embedder for $e(i,j)$ is $\displaystyle \frac{f(i)+f(j)}{2}$, where the edge ends are nodes $i$ and $j$; the node embedding of $i$ and $j$ are $f(i)$ and $f(j)$, respectively.

\end{enumerate}


\textsf{Node2Vec} \cite{grover2016node2vec} can also be used for edge feature representation. The original implementation is found to be slow and memory inefficient \cite{liu2021pecanpy}. Hence, a fast and memory efficient version of \textsf{Node2Vec} called \textsf{PecanPy} (\textbf{P}arallelized, memory \textbf{e}ffi\textbf{c}ient and \textbf{a}ccelerated \textbf{n}ode2vec in \textbf{Py}thon) \cite{liu2021pecanpy, liu2021accurately} is utilized in this paper. \textsf{PecanPy} makes the \textsf{Node2Vec} implementation efficient on the following three fronts:

\begin{itemize}
    \item[(a)] \textbf{Parallelism}: The estimation of transition probability and the random walk generation processes are independent, but are not parallelized in the original \textsf{Node2Vec}. \textsf{PecanPy} parallelizes the walk generation process, which makes the operation much faster.
    \item[(b)] \textbf{Data Structure}: The original implementation of \textsf{Node2Vec} uses NetworkX \cite{hagberg2008exploring} to store graphs which is inefficient for large-scale computation. However, \textsf{PecanPy} uses the Compact Sparse Row (CSR) format for sparse graphs -- which has similar sparsity properties to the transportation network that is addressed in this paper. The CSR formatted graphs are more compactly stored in memory and run faster as they can utilize cache more efficiently.
    \item[(c)] \textbf{Memory}: The original version of \textsf{Node2Vec} pre-processes and stores the 2\textsuperscript{nd} order transition probabilities, which leads to significant memory usage. \textsf{PecanPy} eliminates the pre-processing stage and computes the transition probabilities whenever it is required, without saving.
\end{itemize}

\subsection{Details of GNN}\label{Sec_GNN}

Neural network models for graph-structured data are known as GNNs \cite{chen2018fastgcn, ding2009pagerank, kipf2016semi, velivckovic2017graph}. These models exploit the graph's structure to aggregate the feature information/embeddings of the edges and nodes \cite{hamilton2017inductive}. Feature aggregation from the structured pattern of the graph enables the GNN to predict the probability of edge existence or to predict node labels. The graph structure information is assimilated from the adjacency matrix and the feature information matrix of nodes and edges, which form the inputs, and training using a loss function. Message passing occurs in each GNN layer when each node aggregates the features of its neighbors. The node feature vector is updated by combining its feature vector with the aggregated features from its adjacent nodes. In the first layer, the GNN combines the features of its immediate neighbors, and with an increasing number of layers, the depth of assimilating the neighboring edge features increases accordingly. The edge feature vector is updated with the aggregated features from its adjacent edges, and this procedure repeats for each GNN layer. There are three steps in a GNN, elaborated as follows:
\begin{itemize}

\item \textit{Step 1: Message passing of edge features:}

The original message-passing algorithm is applied to the node features passing via the edges of the graph. Since this work focuses on ranking the edges, each edge is associated with edge features/embedding. Such features are represented using the vector $\mathbb{R}^d$ -- a latent dimensional representation. A popular algorithm to obtain such representation is \textsf{Node2vec} \cite{grover2016node2vec}, previously discussed in Section \ref{Sec_FeatureRep}. The new framework presented in this paper contains a modified version of the message-passing concept -- edge features are aggregated and passed to the neighboring edges. In this way, the GNN learns the structural information. An example of the message passing step is shown in Figure \ref{GNNConcept2}. While conventional implementation of GNNs uses the message passing on the nodes, such passing is performed on the edges here, which is the novelty.

\begin{figure}[h!]
    \begin{center}
        \includegraphics[width=\textwidth]{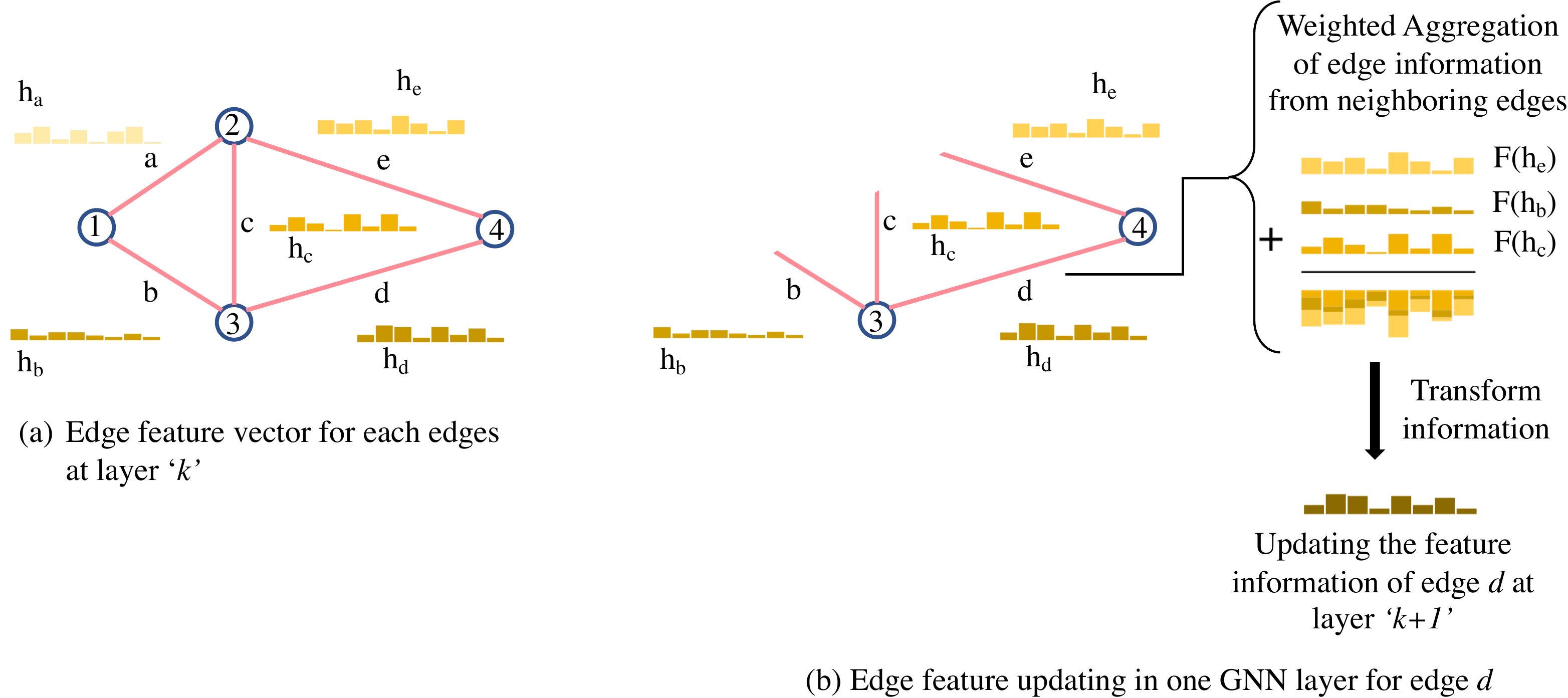}
        \caption{Schematic diagram of the message passing of GNN: (a) This is a sample graph with 4 nodes and 5 edges. Each edge has its own embedding vector shown as $h_i$ for $i$th node, $i \in \{a,b,c,d,e \}$; (b) message passing procedure of edge $d$ using the edge vectors of the neighboring edges $b$, $c$, and $e$ and transforming them, finally ``passing" them to the edge of interest. This process is repeated, in parallel, for all the edges in the graph. The transformation function can be a simple Neural network (RNN or MLP) or an affine transform, $F(h_i) = \textbf{W}_i h_i + b_i$. }
        \label{GNNConcept2}
    \end{center}
\end{figure}

\item \textit{Step 2: Aggregation:}


Messages are aggregated after all the messages from the adjacent edges are passed to the edge of interest. Some popular aggregation functions are:

\begin{equation}\label{Eq_gnn3}
    \mathrm{Sum} = \sum_{j \in \mathcal{N}_i} F(h_j);\quad \mathrm{Mean} = \frac{\displaystyle \sum_{j \in \mathcal{N}_i} F(h_j)}{|{N}_i|}; \quad \mathrm{Max} = \max_{j \in \mathcal{N}_i} F(h_j); \quad \mathrm{Min} = \min_{j \in \mathcal{N}_i} F(h_j)
\end{equation}

\noindent where, $\mathcal{N}_i$ is the set of neighboring edges of edge $i$ (edge of interest). Considering `AGGREGATE' as the aggregation function (sum, mean, max, or min of the neighboring edge message transform), the aggregated message $\mu$ at layer $k$ can be expressed as:

\begin{equation}\label{Eq_gnn4}
    \mu_i^{(k)} = \textrm{AGGREGATE}^{(k)} (\{ h_j^{(k)}: \quad j \in \mathcal{N}_i \})
\end{equation}

\newpage
\item \textit{Step 3: Update:}

These aggregated messages update the source edge's features in the GNN Layer. In this updated step, the edge feature vector is combined with the aggregated messages and is executed by simple addition or concatenation:

\begin{equation}\label{Eq_gnn5}
    \textrm{Addition:}\; h_j^{(k+1)} = \sigma(\Gamma(\Omega(h_j^{(k)})+\mu_i^{(k)})); \qquad \textrm{Concatenation:}\; h_j^{(k+1)} = \sigma(\Gamma(\Omega(h_j^{(k)}) \oplus \mu_i^{(k)}))
\end{equation}

\noindent where, $\sigma$ is the activation function, $\Omega$ is a simple multi-layer perceptron (MLP), and $\Gamma$ is another neural network that projects the added or concatenated vectors to another dimension. In short, the updating step from the previous layer can be summarized as follows:

\begin{equation}\label{Eq_gnn6}
    h_j^{(k+1)} = \textrm{COMBINE}^{(k)} (h_j^{(k)}, \mu_i^{(k)})
\end{equation}

The output of each GNN layer is forwarded as the input to the next GNN layer. After $k$-th GNN layers/iteration, the edge embedding vector at the final layer captures the edge feature information and the graph structure information of all adjacent edges from 1-hop distance to the $k$-th hop distance. The edge feature vector of the 1\textsuperscript{st} layer is obtained using \textsf{NodeVec} \cite{grover2016node2vec} as described in Section \ref{Sec_FeatureRep}.

\end{itemize}

\section{Proposed GNN Framework}\label{Sec_proposedFramework}

Building on the concepts of the GNN presented previously, the proposed GNN framework for estimating approximate edge ranking is presented next. Here, the main reason for choosing GNN over GCN (Graph Convolution Network) \cite{kipf2016semi} is that the aggregated feature vector of a particular edge is dominated by the feature vector of the edge itself as GCN adds a self-loop in the graph. This section introduces the algorithm along with the architecture being proposed. A description of the modified adjacency matrix and its use in edge betweenness ranking is described. This is followed by the details of the edge feature aggregation process in the GNN module. Finally, the details about the ranking loss function are presented.


\subsection{Algorithm and the GNN Architecture}

\begin{figure}[h!]
    \begin{center}
        \includegraphics[width=0.8\textwidth]{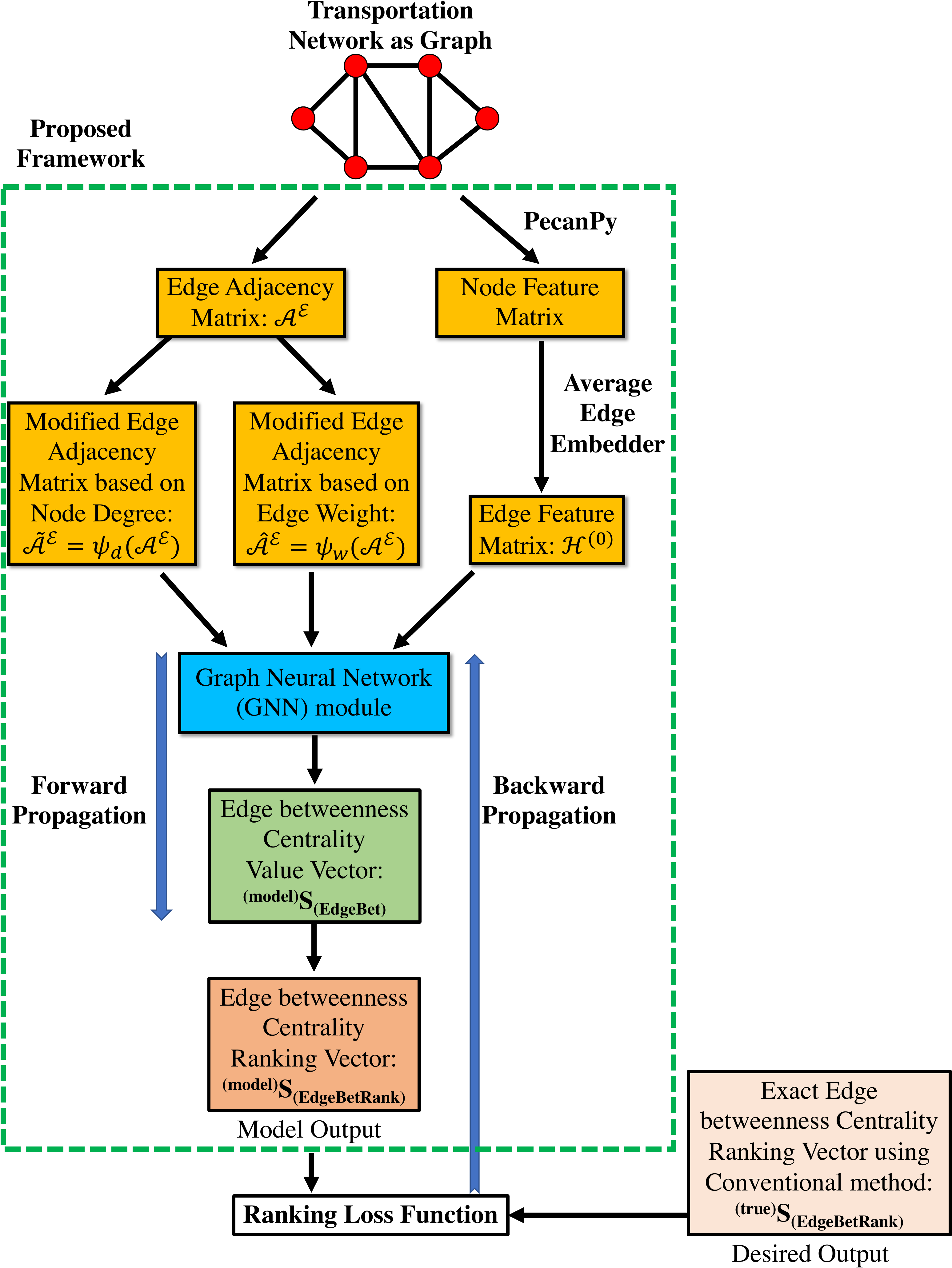}
        \caption{Proposed framework for approximate edge betweenness centrality ranking}
        \label{proposedFrameworkFinal2}
    \end{center}
\end{figure}

Figure \ref{proposedFrameworkFinal2}  shows the overall process of calculating the approximate EBC. This framework takes the graph structure\textemdash specifically the edge adjacency matrix\textemdash and the feature matrix as inputs to estimate the EBC ranking vector depicting the importance of each edge in the graph structure. The GNN module is at the core of this procedure, whose inputs are the edge feature matrix and the two variants of the edge adjacency matrix. Starting with initial weights in the GNN module, the EBC ranking vector of the model is calculated by backpropagating the errors through the GNN layers and then updating the weights iteratively.


\newpage
\subsubsection{Edge Adjacency matrix}\label{Sec_Edge_Adj_mat}
\begin{figure}[h!]
    \begin{center}
        \includegraphics[width=0.7\textwidth]{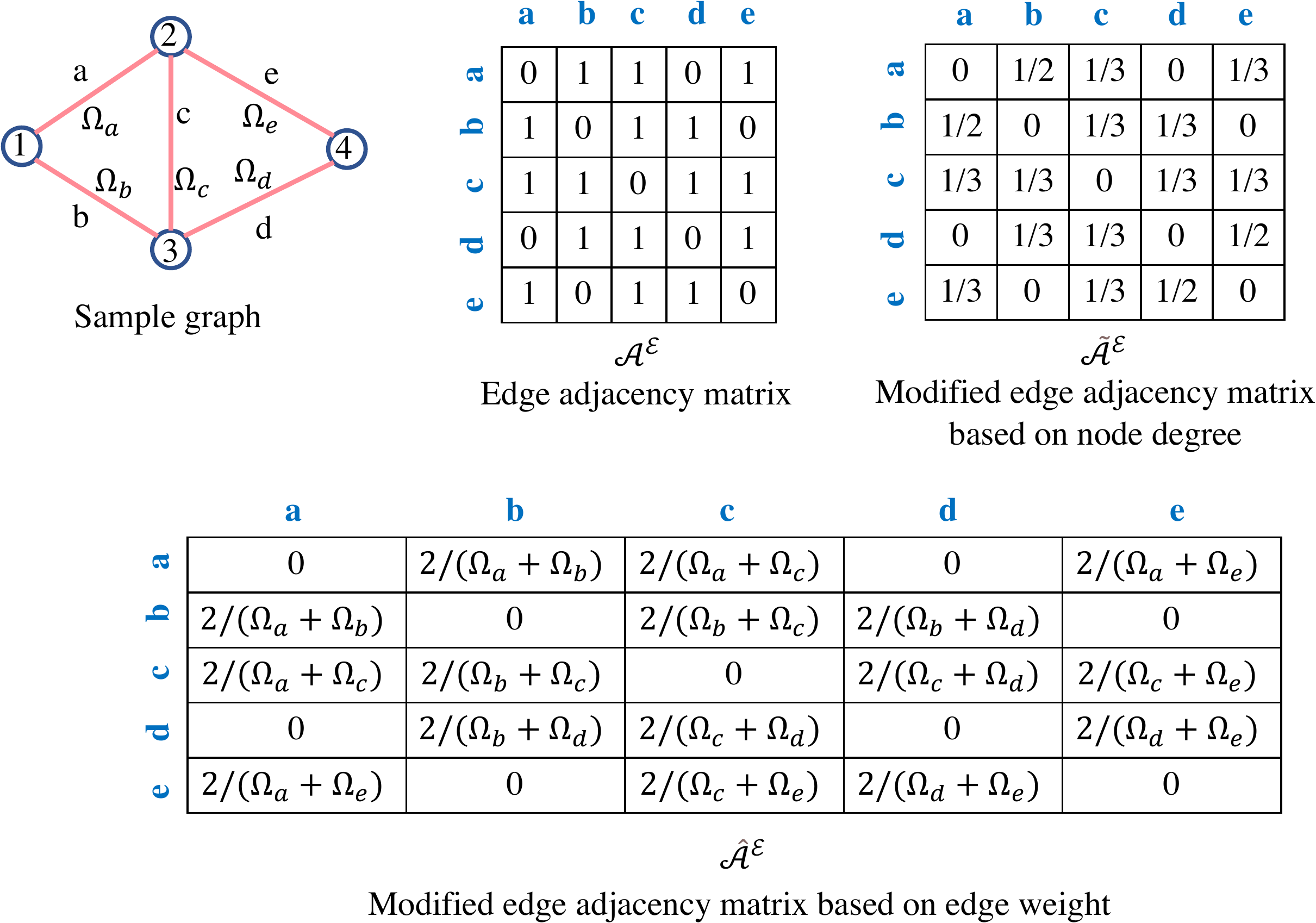}
        \caption{Modified edge adjacency matrices from the edge adjacency matrix of a sample graph network}
        \label{GraphFunctions2}
    \end{center}
\end{figure}

\begin{figure}[h!]
    \begin{center}
        \includegraphics[width=0.9\textwidth]{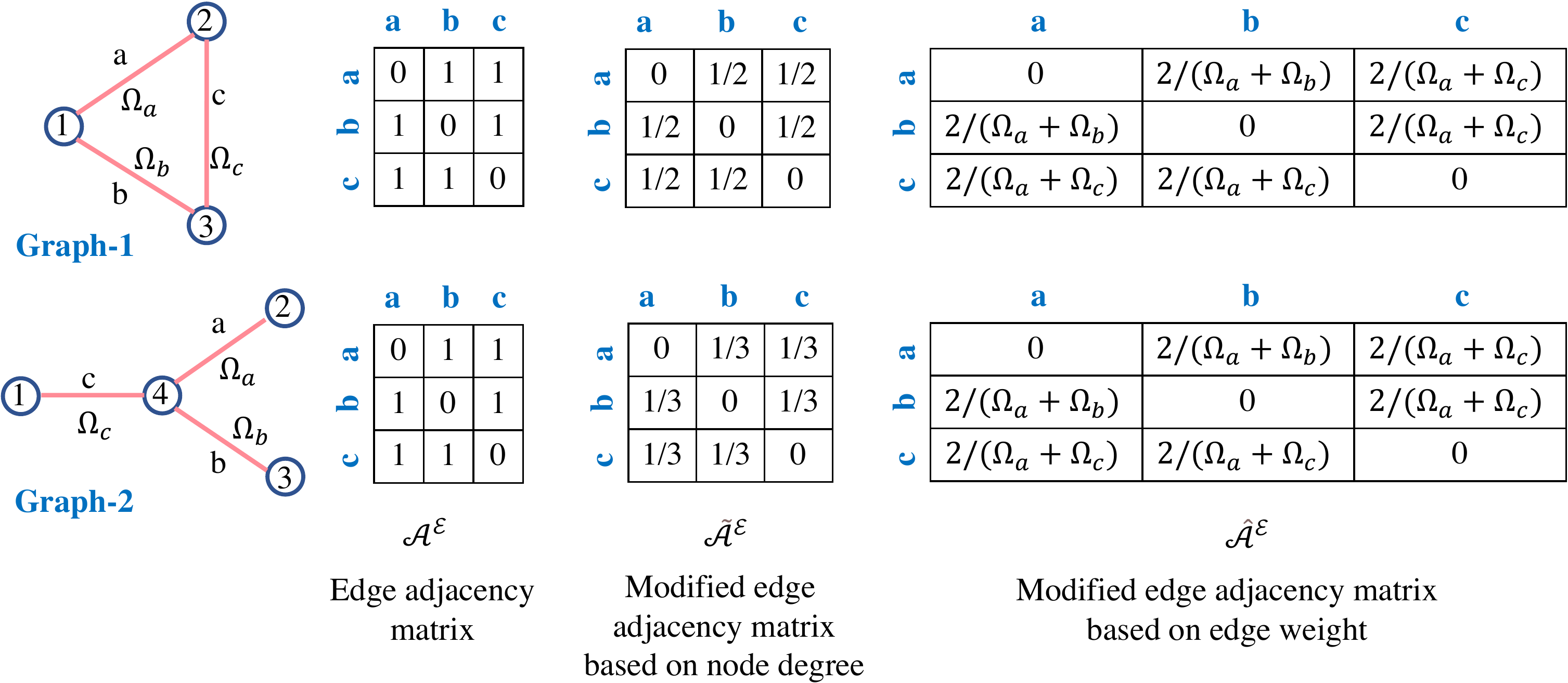}
        \caption{Non-uniqueness of edge adjacency matrix and the corresponding modifications proposed in the new framework}
        \label{EdgeAdjDemoPublish}
    \end{center}
\end{figure}

The edge adjacency matrix is not unique for all graph structures. For instance, a pair of non-isomorphic graphs -- the three point star graph $S_3$ and the cycle graph on three vertices $C_3$ have identical edge-adjacency matrices as shown in Figure \ref{EdgeAdjDemoPublish}. Hence we introduce two variants for this matrix -- modified edge adjacency matrix based on node degree $\Tilde{\mathcal{A}}^{\mathcal{E}}$ and the modified edge adjacency matrix based on edge weight $\Hat{\mathcal{A}}^{\mathcal{E}}$. The modified edge adjacency matrices is obtained from the edge adjacency matrix using the functions $\psi_d$ and $\psi_w$ respectively -- shown in detail in Algorithm \ref{Algo_EdgeBet}: lines 13-32 and the corresponding example is shown in Figure \ref{GraphFunctions2}. The edge weights of edges $a$, $b$, $c$, $d$, and $e$ are $\Omega_a$, $\Omega_b$, $\Omega_c$, $\Omega_d$, and $\Omega_e$, respectively. The matrix $\Tilde{\mathcal{A}}^{\mathcal{E}}$ is unique to each graph; $\Hat{\mathcal{A}}^{\mathcal{E}}$ retains similar features to the original edge adjacency matrix and is non-unique to graph structures. 





\subsubsection{GNN Module}\label{Sec_GNN_module}
\begin{figure}[h!]
    \begin{center}
        \includegraphics[width=0.8\textwidth]{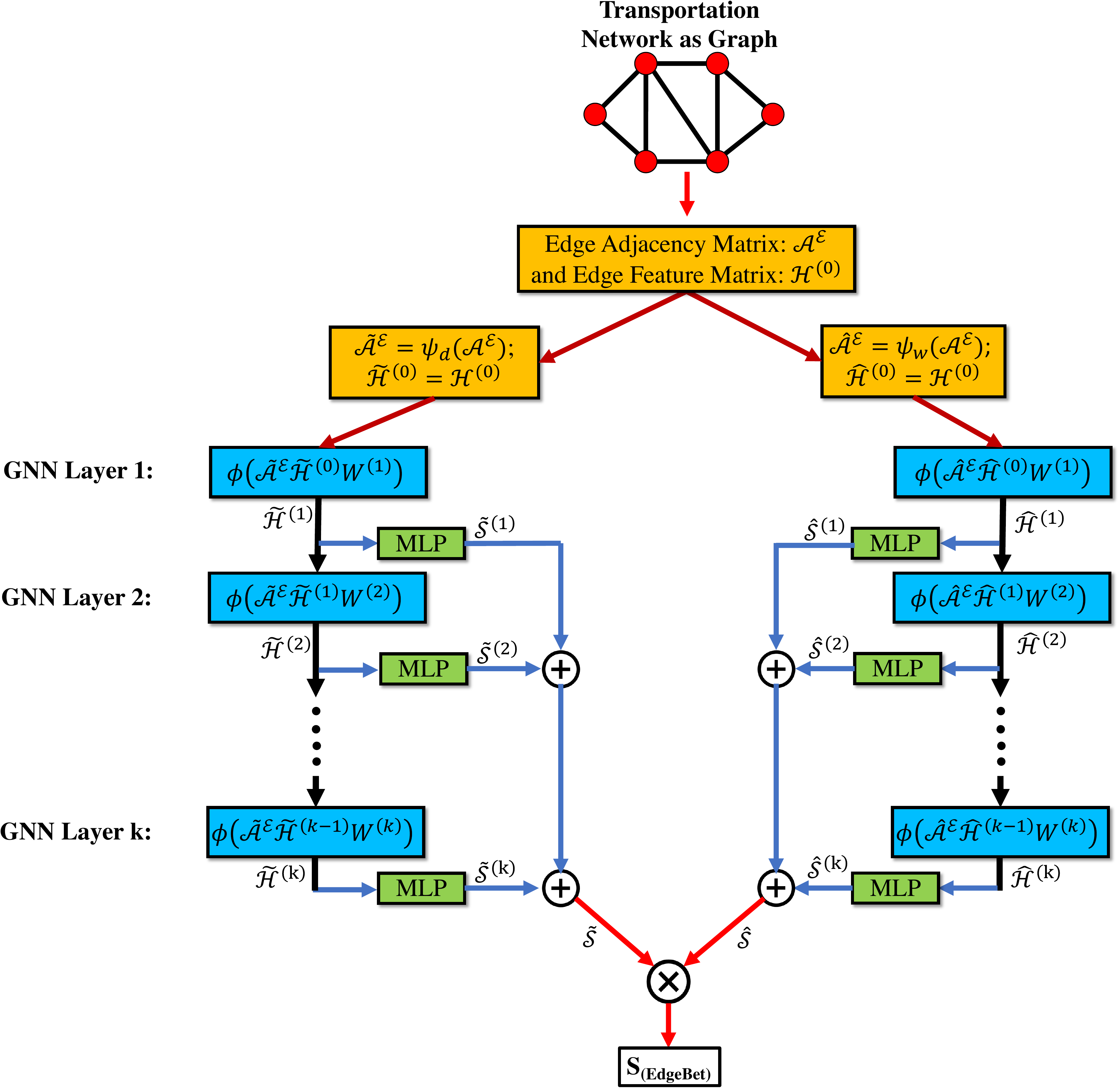}
        \caption{Proposed Graph Neural Network (GNN) architecture. This module takes the edge adjacency matrix and the edge feature/embedding matrix obtained from \textsf{Node2vec/PecanPy} as inputs and calculates the edge importance ranking.}
        \label{GNNArchitecture2}
    \end{center}
\end{figure}

\begin{figure}[h!]
    \begin{center}
        \includegraphics[width=0.3\textwidth]{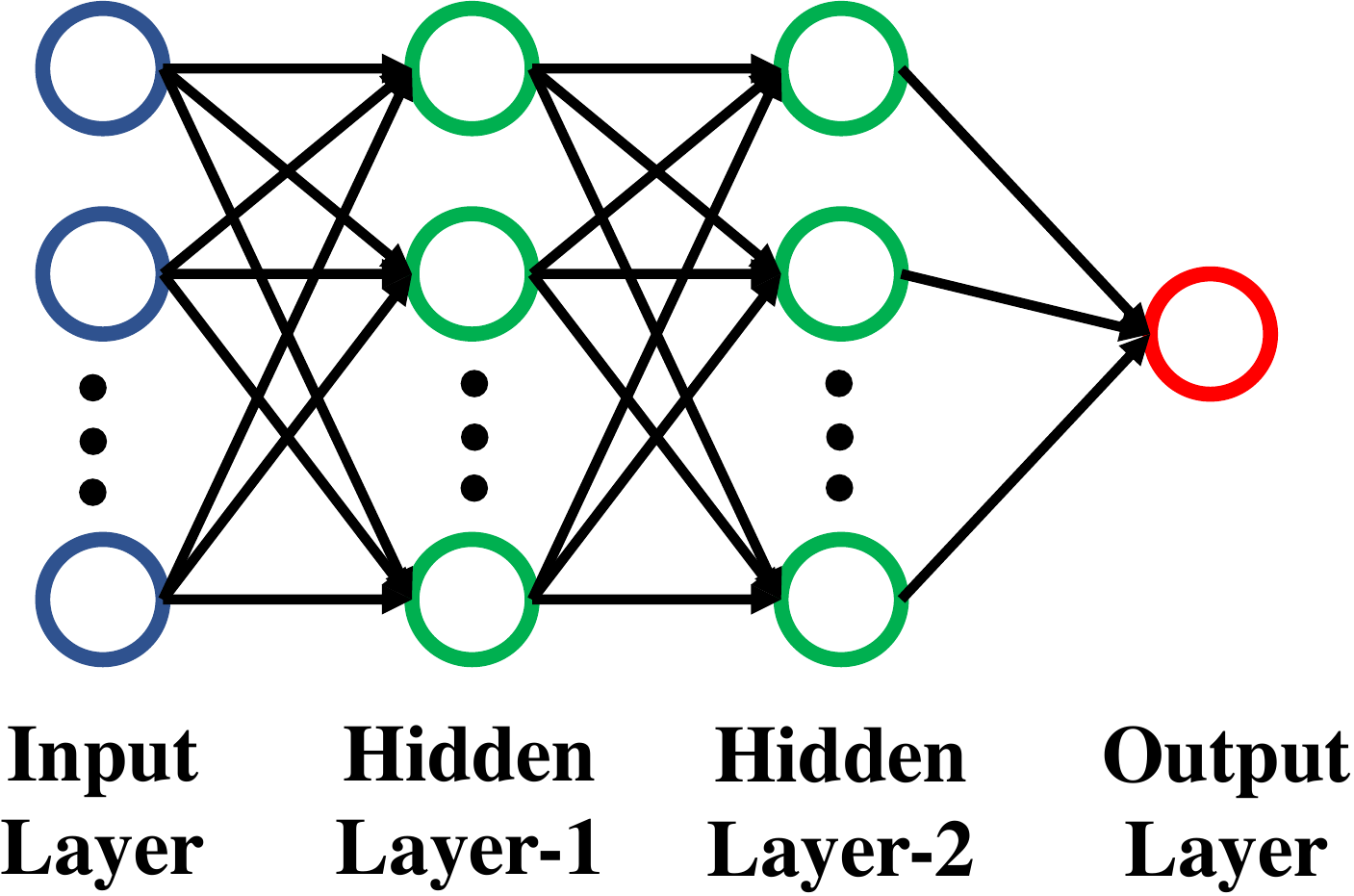}
        \caption{Multi Layer Perceptron (MLP) module}
        \label{MLP2}
    \end{center}
\end{figure}

 The pseudo-code for the GNN framework and the GNN architecture are shown in Algorithm \ref{Algo_EdgeBet} and Figure \ref{GNNArchitecture2}, respectively. The initial feature for an edge is obtained from the edge embeddings as discussed in Section \ref{Sec_FeatureRep}, denoted as $\mathcal{H}^0$. The features of its $k$-hop neighbors are aggregated at the $k$-th layer. A simple summation of the edge feature vectors is used here for aggregation. At each layer, the feature matrix $\mathcal{H}^0$ is multiplied with the modified adjacency matrices i.e., $\Tilde{\mathcal{A}}^{\mathcal{E}}$ and $\Hat{\mathcal{A}}^{\mathcal{E}}$. Then, for each edge, the features of the adjacent edges are summed, as shown in Figure \ref{GNNArchitecture2} and Algorithm \ref{Algo_EdgeBet}: line 5-6, with the Leaky-ReLU activation function. The choice of this activation function is not arbitrary and was an outcome of an extensive exercise with different activation functions (details omitted here for the sake of brevity). Subsequently, the aggregated edge features from each GNN layer are mapped to a Multilayer Perceptron (MLP) unit, which outputs a vector of vector space; $\mathbb{R}^{\mathcal{E}}$, and each value of this vector corresponds to each edge of the network as shown in Figure \ref{MLP2} and Algorithm \ref{Algo_EdgeBet}: line 7-8. During the training phase, the MLP learns to predict a single score based on the input edge features and the graph connection. Single MLP units are implemented in all layers to output the scores, which are then summed separately as $\Tilde{\mathcal{S}}$ and  $\Hat{\mathcal{S}}$ for the modified adjacency matrices $\Tilde{\mathcal{A}}^{\mathcal{E}}$ and $\Hat{\mathcal{A}}^{\mathcal{E}}$, respectively. The MLP unit comprises of three fully connected layers and a hyperbolic tangent as the tuned nonlinearity function. The two scores $\Tilde{\mathcal{S}}$ and $\Hat{\mathcal{S}}$ are multiplied to obtain the final score for each edge as shown in Algorithm \ref{Algo_EdgeBet}: line 12. In this architecture, the weights of all the hidden units are initialized using Xavier initialization \cite{glorot2010understanding} -- which is a standard technique used for weight initialization to ensure that the variance of the activations in every layer is identical. Due to the equal variance in every layer, the exploding or vanishing gradient problems are prevented.  

\begin{algorithm}

\caption{GNN based edge betweenness ranking algorithm (forward propagation)}\label{Algo_EdgeBet}
\textbf{Input:} Number of Edges $\mathcal{E}$, Edge weight List $\Omega$, unweighted Edge adjacency matrix $\mathcal{A}^{\mathcal{E}}$, Feature matrix $\mathcal{H}^0$, GNN depth $K$, GNN weight matrices $W^{(k)}$\\
\textbf{Output:} Edge betweenness centrality value vector $S_{(\textrm{EdgeBet})}$\\

\begin{algorithmic}[1]
\State $ \Tilde{\mathcal{A}}^{\mathcal{E}} \gets  \psi_d (\mathcal{A}^{\mathcal{E}}) $ \Comment{function $\psi_d$ modifies $\mathcal{A}^{\mathcal{E}}$ based on node degree}
\State $ \Hat{\mathcal{A}}^{\mathcal{E}} \gets \psi_w (\mathcal{A}^{\mathcal{E}}) $ \Comment{function $\psi_w$ modifies $\mathcal{A}^{\mathcal{E}}$ based on edge weight}
\State $\Tilde{\mathcal{H}}^{(0)}=\Hat{\mathcal{H}}^{(0)}=\mathcal{H}^0$ \vspace{5mm}
\For{$k = 1, \cdots, K$}
    \State $\Tilde{\mathcal{H}}^{(k)} \gets \phi (\Tilde{\mathcal{A}}^{\mathcal{E}} \; \Tilde{\mathcal{H}}^{(k-1)} \; W^{(k)} )$
    \State $\Hat{\mathcal{H}}^{(k)} \gets \phi (\Hat{\mathcal{A}}^{\mathcal{E}} \; \Hat{\mathcal{H}}^{(k-1)} \; W^{(k)} )$  \Comment{$\phi$ is the activation function}
    \State $\Tilde{\mathcal{S}}^{(k)} \gets \textrm{MLP} (\Tilde{\mathcal{H}}^{(k)})$
    \State $\Hat{\mathcal{S}}^{(k)} \gets \textrm{MLP} (\Hat{\mathcal{H}}^{(k)})$ \Comment{MLP is the multi-layer perceptron}
\EndFor \vspace{5mm}
\State $\Tilde{\mathcal{S}} \gets \displaystyle \sum_{k = 1, \cdots, K} |\Tilde{\mathcal{S}}^{(k)}| $
\State $\Hat{\mathcal{S}} \gets \displaystyle \sum_{k = 1, \cdots, K} |\Hat{\mathcal{S}}^{(k)}| $
\State $S_{(\textrm{EdgeBet})} \gets \Tilde{\mathcal{S}} \times \Hat{\mathcal{S}}$ \vspace{5mm}

\Function{$\psi_d$}{$\mathcal{A}^{\mathcal{E}}$}
    \State $ \Tilde{\mathcal{A}}^{\mathcal{E}} = zeros(\mathcal{E}, \mathcal{E})$
    \For{$i = 1, \cdots, \mathcal{E}$}
        \For{$j = i+1, \cdots, \mathcal{E}$}
            \State $ \Tilde{\mathcal{A}}^{\mathcal{E}}(i,j)=\displaystyle\frac{\mathcal{A}^{\mathcal{E}}(i,j)}{\textrm{Degree of the node connecting edge $i$ and $j$}}$
            \State $ \Tilde{\mathcal{A}}^{\mathcal{E}}(j,i)=\Tilde{\mathcal{A}}^{\mathcal{E}}(i,j)$
        \EndFor
    \EndFor
    \State \Return $ \Tilde{\mathcal{A}}^{\mathcal{E}}$
\EndFunction \vspace{5mm}

\Function{$\psi_w$}{$\mathcal{A}^{\mathcal{E}}$}
    \State $ \Hat{\mathcal{A}}^{\mathcal{E}} = zeros(\mathcal{E}, \mathcal{E})$
    \For{$i = 1, \cdots, \mathcal{E}$}
        \For{$j = i+1, \cdots, \mathcal{E}$}
            \State $ \Hat{\mathcal{A}}^{\mathcal{E}}(i,j)=\displaystyle\frac{\mathcal{A}^{\mathcal{E}}(i,j)}{(\Omega_i+\Omega_j)/2}$
            \State $ \Hat{\mathcal{A}}^{\mathcal{E}}(j,i)=\Hat{\mathcal{A}}^{\mathcal{E}}(i,j)$
        \EndFor
    \EndFor
    \State \Return $ \Hat{\mathcal{A}}^{\mathcal{E}}$
\EndFunction

\end{algorithmic}
\end{algorithm}

\clearpage

\subsection{Loss Function}\label{Sec_Loss_Function}

A ranking loss function is used to estimate the loss due to the differences in the ranking predicted by the proposed model compared to the target EBC ranking as presented in Section \ref{Sec_EBC_Conventional}. Such ranking loss functions have previously been used for recommendation systems to rank or rate products or users \cite{burges2010ranknet, chen2009ranking}. The margin ranking loss function is defined as follows:

\begin{align}
    \mathscr{L} (S_i^{(\textrm{model})}, S_i^{(\textrm{true})}, y) &= \max (0, -y \cdot (S_i^{(\textrm{model})}-S_i^{(\textrm{true})}) + \textrm{Margin}) \\
y &=  \begin{cases}
1, \qquad \textrm{if} \; S_i^{(\textrm{model})} \; \textrm{should be ranked higher than} \; S_i^{(\textrm{true})} \\
-1, \qquad \textrm{if} \; S_i^{(\textrm{true})} \; \textrm{should be ranked higher than} \;  S_i^{(\textrm{model})} \nonumber
\end{cases}
\end{align}

\noindent where, $S_i^{(model)}$ is the predicted ranking-score, and $S_i^{(true)}$ is the EBC score obtained using the conventional method i.e., Brandes Algorithm \cite{brandes2008variants} as shown in Equation \ref{Eq_3}. In this study, the margin value is set to 1 to allow some flexibility. 


\section{Results}\label{Sec_Results}

\noindent Experimental results for both the synthetic and real-world cases are presented in this section. First, details of training for the proposed architecture are discussed. Then, the network performance on synthetic graphs and actual transportation network data are presented. 


In terms of computing resources, the experiments were conducted on a dedicated computer and the hardware and the software information for the computing resources used are shown in Table \ref{Tab_HardwareInfo}.

\begin{table}[h!]
\centering
\caption{Hardware and software information for the deep learning framework}
\begin{tabular}{|l|l|} \hline
CPU model and speed         & Intel Core i9-10940X CPU @ 3.30 GHz \\ 
Available Hard disk         & 1 TB                     \\
Available RAM               & 64 GB                     \\
GPU type and specification           & NVIDIA GeForce RTX 3090 -- 32 GB          \\
Programming                 & Python 3.7, Matlab 2022a          \\
Deep Learning framework     & PyTorch, Numpy, Scipy, CUDA 11.6 \\ 
GNN framework     & NetworkX, \textsf{Node2Vec}, \textsf{Pecanpy}, Gensim   \\  \hline
\end{tabular}\label{Tab_HardwareInfo}
\end{table}

The graph datasets are divided into training ($\approx$90\%) and testing ($\approx$10\%) datasets -- which is typical in machine learning literature \cite{nguyen2021influence}. The EBC ranking is calculated using the conventional method for all graphs in the training and testing datasets. These rankings are used as target vectors for training the GNN. The test graphs are not used for training; the model learns to map edge features and importance via MLP to the ranking scores. Training and testing graphs contain variable numbers of nodes and edges, and the GNN is trained and tested on the same type of synthetic graph. The evaluation metrics used are Kendall's Tau rank correlation coefficient and the Spearman's rank correlation coefficient (details are provided in \ref{EvalMets}).

The model size is determined by the graph, precisely by the number of edges contained in the largest graph. A model size of 10000 is used here to accommodate a typical medium-sized urban area in the US. While the size of the edge adjacency matrix (input) size is fixed at 10000 edges (Figure \ref{GNNArchitecture2}), smaller graphs can also be accommodated by populating only the upper left portion of this matrix, with remaining elements of this input matrix set to zeros (zero-padding). 
The network is trained using the ADAM (Adaptive Moment Estimation) \cite{kingma2014adam} optimizer, which is a variant of stochastic gradient descent (SGD) algorithm commonly used in deep learning. The training hyper-parameters was performed with a learning rate of 0.0005 and a dropout ratio of 0.3. The number of epochs used for training is 50 and the number of hidden neurons (embedding dimension) and the number of GNN layers (\ref{Sec_Ablation}) were optimized. The experiments use an embedding dimension of 256 and 5 layers. The edge features are obtained using \textsf{PecanPy} as discussed in Section \ref{Sec_FeatureRep} with the feature vector of length 256. BFS approach with $p=1$ and $q=2$ is used to search the shortest path. The calculation of the ranking loss function requires a comparison of edge pair rankings. However, ranking all possible combinations of edge pairs -- $|\mathcal{E}| \choose 2$ for $|\mathcal{E}|$ edges is cumbersome, and so the number of edge pairs are randomly sampled as 20 times the number of edges -- which is a common practice \cite{maurya2021graph}.



\subsection{Performance on synthetic networks}

First, three synthetic random graphs are used to evaluate the performance of the proposed method: (a) Erd{\H{o}}s - R{\'e}nyi variant-I \cite{erdHos1960evolution} i.e., $G_{np}$, which is an undirected graph containing $n$ nodes (fixed number) where each edge $(u,v)$ appears independent and identically distributed with probability $p$; (b) Erd{\H{o}}s - R{\'e}nyi variant-II \cite{erdHos1960evolution} i.e., $G_{nm}$, which is an undirected graph containing $n$ nodes (fixed number) and $m$ edges. Edges are connected uniformly to random nodes. Unlike Erd{\H{o}}s - R{\'e}nyi variant-I, the number of edges in Erd{\H{o}}s - R{\'e}nyi variant-II are fixed. Both variants have a fixed number of nodes; and (c) Watts–Strogatz model \cite{watts1998collective} which is a random graph generation model that produces graphs with small-world properties such as local clustering and average shortest path lengths. The small world random graph has been used in applications such as electric power grids, networks of brain neurons, airport networks, etc. \cite{newman2000models}.

\subsubsection{Graph generation parameters}
The synthetic graph generation parameters are shown in Table \ref{Tab_SyntheticGraph_GenParam}. Here, $\mathcal{U}\{a,b\}$ and $\mathcal{U}[a,b]$ represent discrete and continuous uniform distributions between the ranges $a$ and $b$, respectively. With the graph generation parameters chosen arbitrarily for this case study, we generate 1000 training graphs, 100 validation graphs, and 100 test graphs for each case. 
\begin{table}[h!]
\centering
\caption{Generation parameters of the Synthetic Graphs}
\begin{tabular}{|c|c|c|} \hline
Synthetic Graph   Type                & \multicolumn{2}{c|}{Generation Parameters}                        \\ \hline
\multirow{3}{*}{Erd{\H{o}}s - R{\'e}nyi-I (GNP)}  & Nos of nodes                 & $\mathcal{U}\{1000,5000\}$                  \\
                                      & Probability of edge creation & 1.2/(Nos of nodes -1)             \\ 
                                      & Edge weights & $\mathcal{U}[0,100]$             \\ \hline
\multirow{3}{*}{Erd{\H{o}}s - R{\'e}nyi-II (GNM)} & Nos of nodes                 & $\mathcal{U}\{1000,5000\}$                 \\
                                      & Nos of edges                 & $\mathcal{U}[1.4,1.6]\times$  Nos of nodes \\ 
                                      & Edge weights & $\mathcal{U}[0,100]$             \\ \hline
\multirow{3}{*}{Watts-Strogatz}       & Nos of nodes                 & $\mathcal{U}\{2000,4000\}$                  \\
                                      & Mean degree                  & 4                                 \\
                                      & Probability of edge rewiring & 0.5   \\ 
                                      & Edge weights & $\mathcal{U}[0,100]$             \\ \hline
\end{tabular}\label{Tab_SyntheticGraph_GenParam}
\end{table}
Table \ref{Tab_SyntheticGraphStats} summarizes the results generated from the synthetic graphs. The average shortest path length is defined as:

\begin{equation}
    a = \sum_{s,t \in \mathcal{V}} \frac{d(s,t)}{|\mathcal{V}| \cdot (|\mathcal{V}| -1)}
\end{equation}

\noindent where, $\mathcal{V}$ is the set of nodes in weighted/unweighted graph $\mathcal{G}$ of total node number $|\mathcal{V}|$ (cardinality) and $d(s,t)$ is the shortest path from $s$ to $t$. The clustering coefficient is the measure for the nodes that are clustered together, which is the geometric average of the subgraph edge weights \cite{onnela2005intensity}:

\begin{equation}
    c_s = \frac{\displaystyle\sum_{tu} (\hat{w}_{st} \hat{w}_{tu} \hat{w}_{us})^{1/3}}{\mathrm{deg}(s)(\mathrm{deg}(s)-1)}; \qquad C = \frac{1}{|\mathcal{V}|} \sum_{s \in \mathcal{V}} c_s
\end{equation}

\noindent where, $c_s$ is the clustering coefficient of node $s$, $C$ is the average clustering coefficient of the graph; $\mathrm{deg}(s)$ is the degree of node $s$; nodes $s$, $t$, and $u$ create triangles in the graph. The edge weights $\hat{w}_{st}$ are normalized by the maximum weight in the network, $\hat{w}_{st}=w_{st}/\max(w)$.

\begin{table}[h!]
\centering
\caption{Statistics of the synthetic graphs used for the training and testing the proposed GNN framework}
\begin{tabular}{|c|ccc|} \hline
Synthetic Graph              & Erd{\H{o}}s - R{\'e}nyi - I                     & Erd{\H{o}}s - R{\'e}nyi - II                    & Small World Network               \\
Types              & GNP random                     & GNM random                    & Watts-Strogatz Model               \\ \hline
Number of Nodes              & \multirow{2}{*}{Upto 5000}                           & \multirow{2}{*}{Upto 5000}                           & \multirow{2}{*}{Upto 5000}                         \\
(Range)             &                                     &                                     &                                   \\
Number of Edges             & \multirow{2}{*}{Upto 10000}                          & \multirow{2}{*}{Upto 10000}                          & \multirow{2}{*}{Upto 10000}                        \\
(Range)            &                                     &                                     &                                   \\
Avg. Shortest Path Lengths              & \multirow{2}{*}{$277.5 \pm 15.7$, 307.5} & \multirow{2}{*}{$290.6 \pm 20.6$, 338.6} & \multirow{2}{*}{$249.4 \pm 7.4$, 266.6}  \\
 ($\mu \pm \sigma$, max)             &                                     &                                     &                                   \\
Avg. Clustering Coeff.              & $9.04\times10^{-4} $    & $1.02\times10^{-4} $    & $0.0681$  \\
($\mu \pm \sigma$)                  &      $\pm7.91\times10^{-4}$                               &    $\pm7.97\times10^{-4}$                                 &        $\pm0.004$                           \\
Average Degree of             & \multirow{2}{*}{$3.198 \pm 0.031$}       & \multirow{2}{*}{$3.182 \pm 0.102$}       & \multirow{2}{*}{$4\pm0$}        \\
Nodes ($\mu \pm \sigma$)                  &                                     &                                     &                      \\ \hline            
\end{tabular}\label{Tab_SyntheticGraphStats}
\end{table}

\subsubsection{Training time of the model and speed for inference (latency)}

\begin{figure}[h!]
    \begin{center}
        \includegraphics[width=\textwidth]{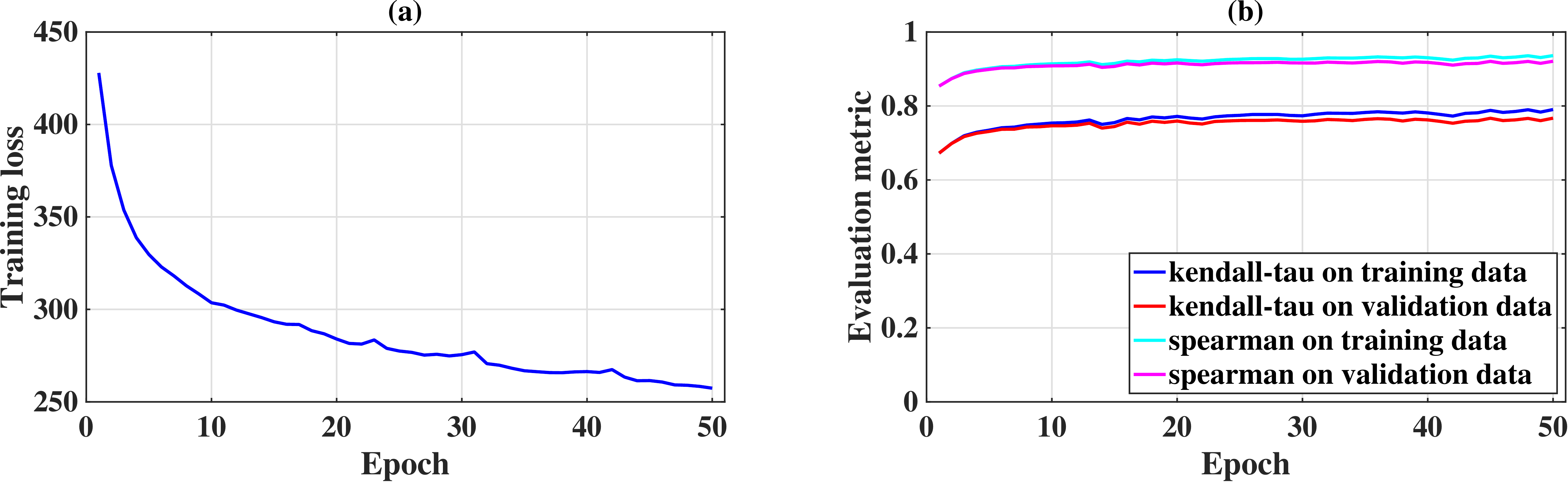}
        \caption{Evolution in learning phase -- (a) Training loss over epochs and (b) Evaluation metric (for training and validation data) over epochs.}
        \label{LossAccuVsEpoch}
    \end{center}
\end{figure}

Figure \ref{LossAccuVsEpoch}(a) shows the evaluated training marginal ranking loss evolution for every epoch. The ranking scores (Kendall's-Tau and Spearman's correlation) for training and validation data are also shown in Figure \ref{LossAccuVsEpoch}(b). Given the asymptotic nature of the evaluation metric, the training was stopped at 50 epochs. Each epoch took approximately 175 seconds to train on average --  hence the total training time for the graphs containing 10000 edges took $\approx 2.43$ hours.




\begin{figure}[h!]
    \begin{center}
        \includegraphics[width=0.7\textwidth]{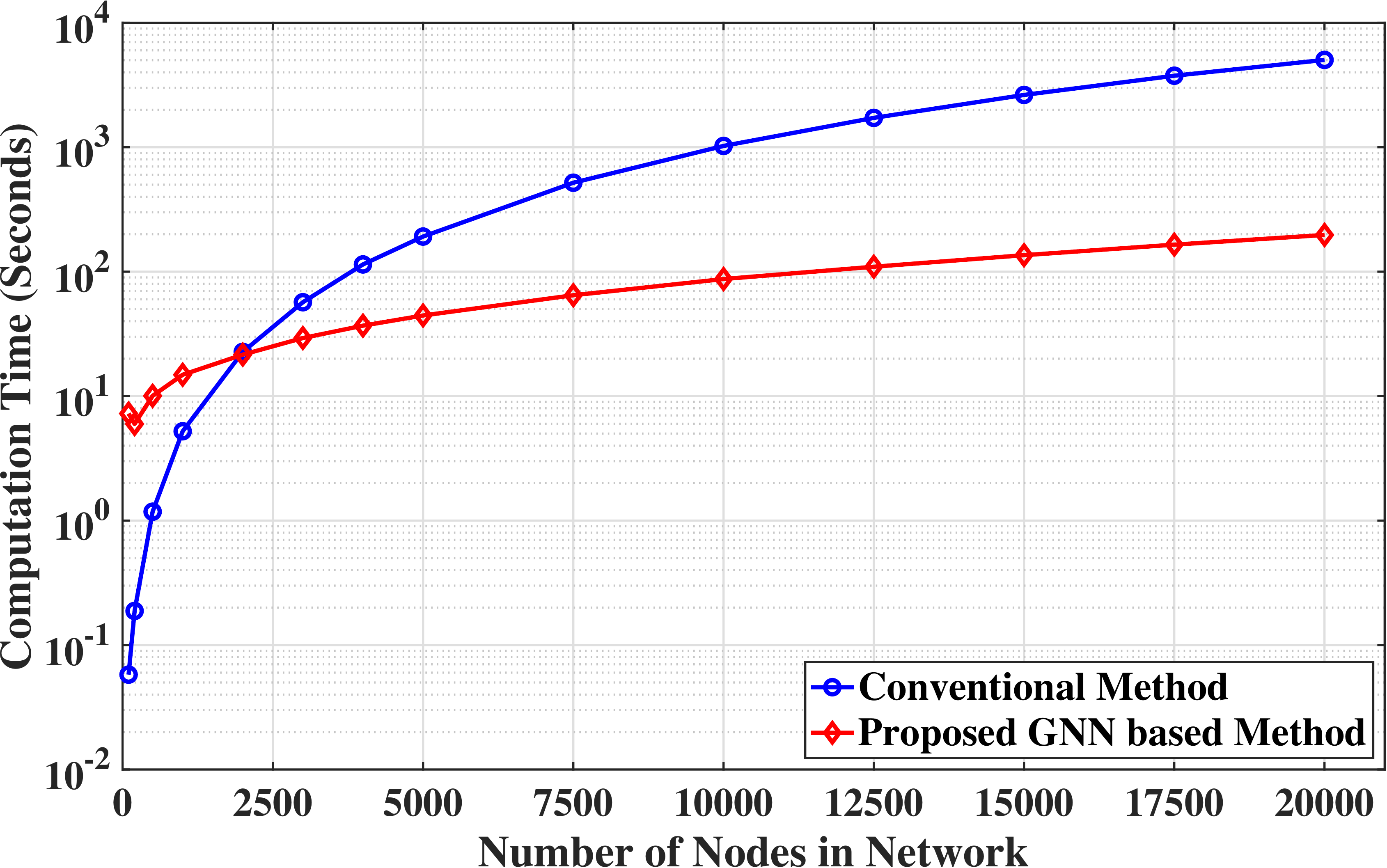}
        \caption{Comparison of computing time for edge ranking of graphs between the conventional method and the proposed GNN based approach}
        \label{testingTimePlot2}
    \end{center}
\end{figure}

The inference speed combines the latencies associated with GNN and \textsf{PecanPy}. While the computational overhead with the inference part of the GNN is of milliseconds, \textsf{PecanPy} has a relatively large computational overhead. Figure \ref{testingTimePlot2} shows the comparison of the results between the proposed GNN-based approach and the conventional method -- Brandes' \cite{brandes2001faster} in a semi-log plot. Beyond a graph size of about 2000 nodes, the proposed GNN method outperforms the conventional method. For a graph with 20000 nodes, the traditional process takes approximately 5033 seconds to compute, while the GNN takes a fraction of that time, 197 seconds. These results underscore the advantage of the proposed GNN method for large graphs compared to the conventional method.


\subsubsection{Ranking scores}

\begin{figure}[h!]
    \begin{center}
        \includegraphics[width=\textwidth]{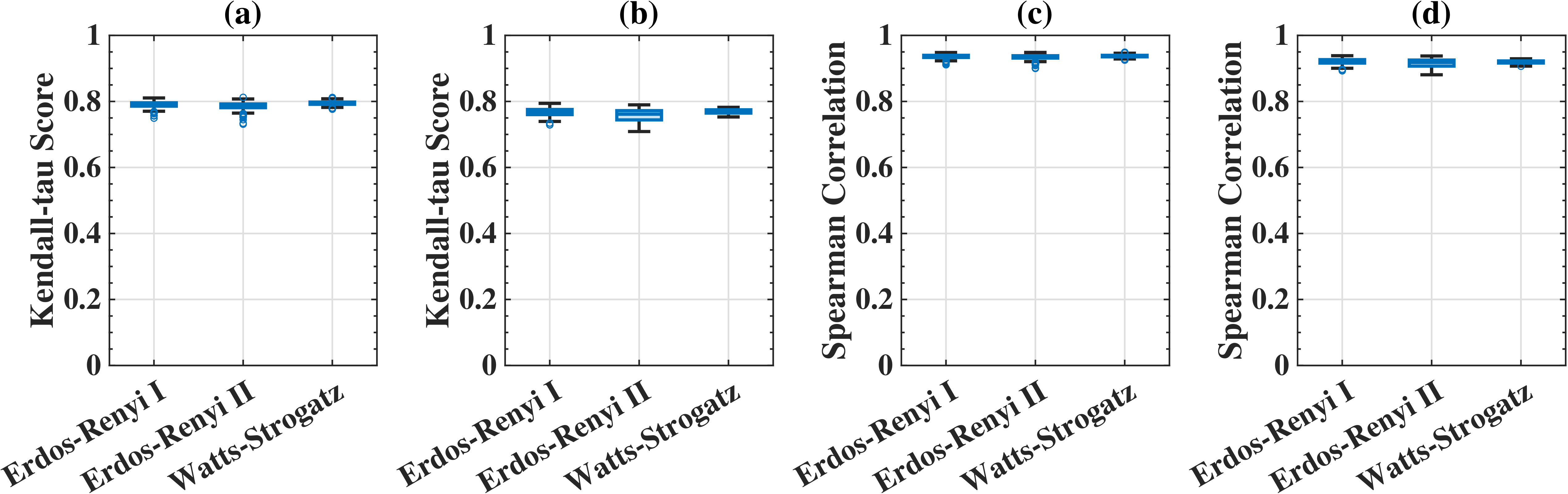}
        \caption{The training and testing ranking score distributions for synthetic graphs; the Box and Whisker plot shows the median, the lower and upper quartiles, any outliers (calculated using the interquartile range), and the minimum and maximum values that are not outliers -- (a) Kendall-tau score on training data, (b) Kendall-tau score on testing data, (c) Spearman correlation coefficient on training data, (d) Spearman correlation coefficient on testing data.}
        \label{SyntheticPerformance}
    \end{center}
\end{figure}

The evaluation metrics (Kendall-Tau \cite{kendall1938new} and Spearman correlation \cite{myers2004spearman}) are calculated for both training and testing data sets for all three types of synthetic graphs and are shown in Tables \ref{Tab_SyntheticGraph_training} and \ref{Tab_SyntheticGraph_testing}, respectively. Details about these evaluation metrics are discussed in \ref{EvalMets}. The ranking scores are sufficiently high, indicating the proposed framework can predict the target values very well. The detailed ranking score statistics in the form of the Box and Whisker plot are also shown in Figure \ref{SyntheticPerformance}. The standard deviations for the estimated scores are minimal, denoting the robustness of the proposed method even though the graph sizes differ substantially in the testing dataset.



\begin{table}[h!]
\centering
\caption{Ranking Scores in Synthetic Graphs (Training Data)}
\begin{tabular}{|c|ccc|} \hline
Synthetic Graph            & Erd{\H{o}}s - R{\'e}nyi - I               & Erd{\H{o}}s - R{\'e}nyi - II              & Small World Network          \\
Types                      & GNP random                    & GNM random                    & Watts-Strogatz Model         \\ \hline
Kendall Tau Score          & \multirow{2}{*}{$0.791\pm0.004$} & \multirow{2}{*}{$0.786\pm0.009$}  & \multirow{2}{*}{$0.795\pm0.005$} \\
($\mu \pm \sigma$) &                               &                               &                              \\
Spearman's Rho Score       & \multirow{2}{*}{$0.936\pm0.005$} & \multirow{2}{*}{$0.934\pm0.006$} & \multirow{2}{*}{$0.938\pm0.003$} \\
($\mu \pm \sigma$) &                               &                               &                             \\ \hline 
\end{tabular}\label{Tab_SyntheticGraph_training}
\end{table}

\begin{table}[h!]
\centering
\caption{Ranking Scores in Synthetic Graphs (Testing Data)}
\begin{tabular}{|c|ccc|} \hline
Synthetic Graph            & Erd{\H{o}}s - R{\'e}nyi - I               & Erd{\H{o}}s - R{\'e}nyi - II              & Small World Network          \\
Types                      & GNP random                    & GNM random                    & Watts-Strogatz Model         \\ \hline
Kendall Tau Score          & \multirow{2}{*}{$0.7670\pm0.013$} & \multirow{2}{*}{$0.758\pm0.017$}  & \multirow{2}{*}{$0.770\pm0.006$} \\
($\mu \pm \sigma$) &                               &                               &                              \\
Spearman's Rho Score       & \multirow{2}{*}{$0.921\pm0.009$} & \multirow{2}{*}{$0.916\pm0.012$} & \multirow{2}{*}{$0.920\pm0.005$} \\
($\mu \pm \sigma$) &                               &                               &                             \\ \hline 
\end{tabular}\label{Tab_SyntheticGraph_testing}
\end{table}


\newpage
\subsubsection{Comparison with other variants of edge adjacency matrix}

Two variants of edge-adjacency matrices have been used in the proposed framework, as shown in Figure \ref{GNNArchitecture2}. This section shows the comparison of the proposed framework with other cases where only one variant of edge adjacency matrix is used for Erd{\H{o}}s-R{\'e}nyi-II type graph networks. Table \ref{Tab_AdjMatComparison} shows that combining $\Tilde{\mathcal{A}}^{\mathcal{E}}$ and $\Hat{\mathcal{A}}^{\mathcal{E}}$ produces better outcomes (scores) than using the edge adjacency matrix, or the modified edge adjacency matrices, individually.

\begin{table}[h!]
\centering
\caption{Ranking Scores in GNM for different adjacency matrix (Testing Data)}
\begin{tabular}{|c|cc|} \hline
Type of adjacency matrix as  & Kendall Tau & Spearman's Rho \\ \hline
$\mathcal{A}^{\mathcal{E}} $                   & 0.329       & 0.476     \\
 $\Tilde{\mathcal{A}}^{\mathcal{E}} $             & 0.339       & 0.489     \\
$\Hat{\mathcal{A}}^{\mathcal{E}}  $             & 0.745       & 0.908     \\
Both $\Tilde{\mathcal{A}}^{\mathcal{E}}$   and $\Hat{\mathcal{A}}^{\mathcal{E}}$     & 0.759       & 0.916    \\ \hline 
\end{tabular}\label{Tab_AdjMatComparison}
\end{table}



\subsection{Application to the Minnesota transportation network}
\begin{figure}[h!]
    \begin{center}
        \includegraphics[width=0.8\textwidth]{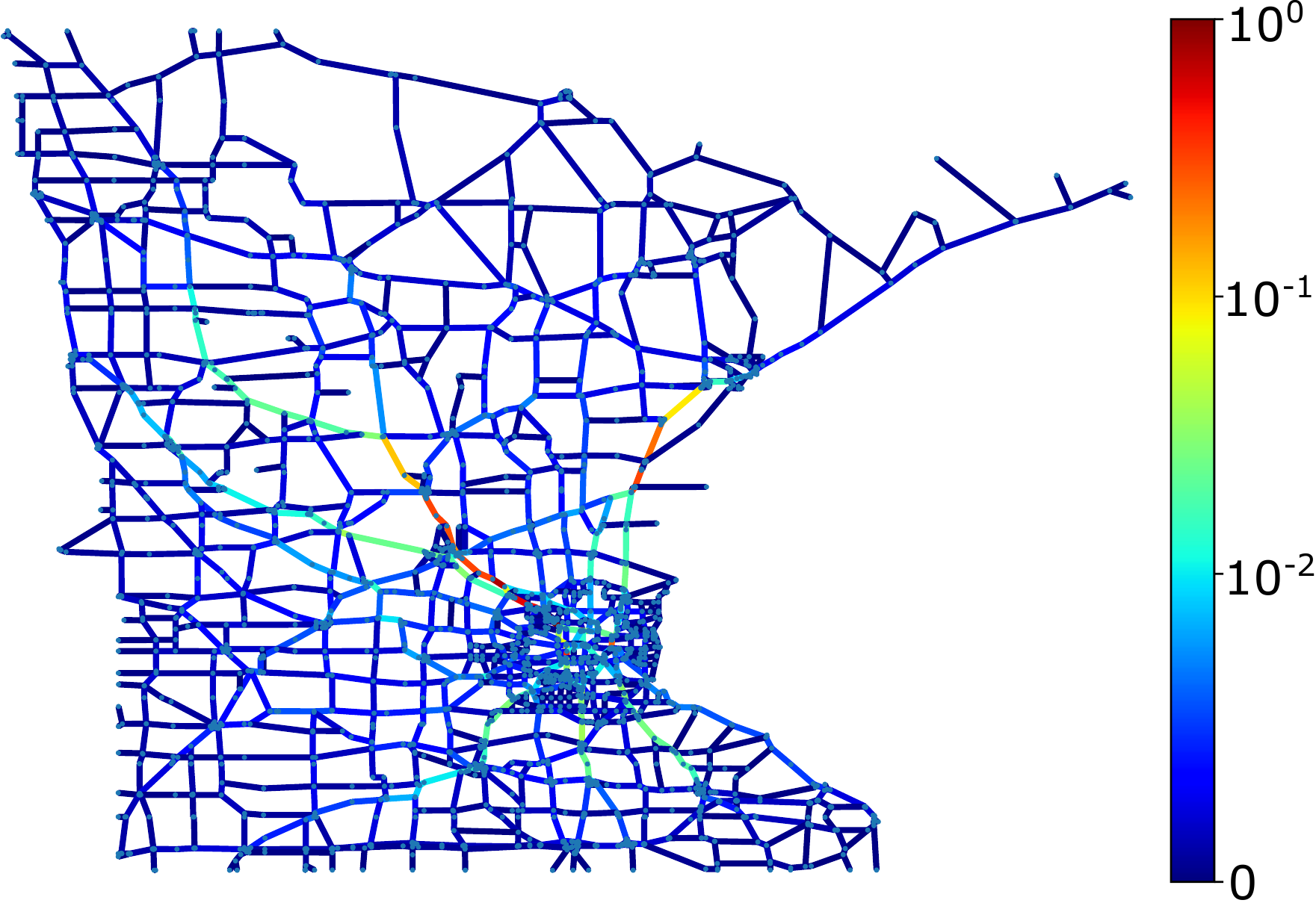}
        \caption{Edge betweenness centrality score for Minnesota State transportation network}
        \label{minebc2}
    \end{center}
\end{figure}

We validated the GNN-based framework on the transportation network for the state of Minnesota, USA. The network information is obtained from the network repository \cite{nr2015}, which contains 2640 nodes (road junctions) and 3302 edges (streets). This network offers a good balance of size and computational overhead to demonstrate the proposed framework's performance aspects. Figure \ref{minebc2} shows the EBC scores for the network. In this Figure, the streets highlighted in red denote the most critical roads, i.e., bridges, in the graph as determined by the EBC measure -- any modification to these streets (change in edge weights, addition, and deletion of roads and junctions) will impact the network significantly. This study simulates the edge importance ranking due to a dynamic change in the parameters of the network -- such as new road construction or inoperative roads, or change in other factors such as travel time, travel distance, or traffic flow.

The experiments consider two cases: (a) change in edge weights, say arising due to varying traffic volumes, and (b) change in edge weights coupled with a permanent change (deletion or addition) of edges, say due to different traffic volumes combined with some inoperative roads or the addition of new roads. In both these cases, the edge weights are determined from the Euclidean distance (2-norm) between the node coordinates, i.e., the road segment length. These cases simulate post-catastrophic events such as a major earthquake or flood, where the first case could be pre-event or after minor events where all the streets are operating. Still, their properties have changed. The latter may result from a significant event where some segments are taken out of service, and the remaining are functional with significantly modified properties.

\subsubsection{Case I: Change in edge weights}\label{Sec_minnesota_case1}
\begin{figure}[h!]
    \begin{center}
        \includegraphics[width=0.7\textwidth]{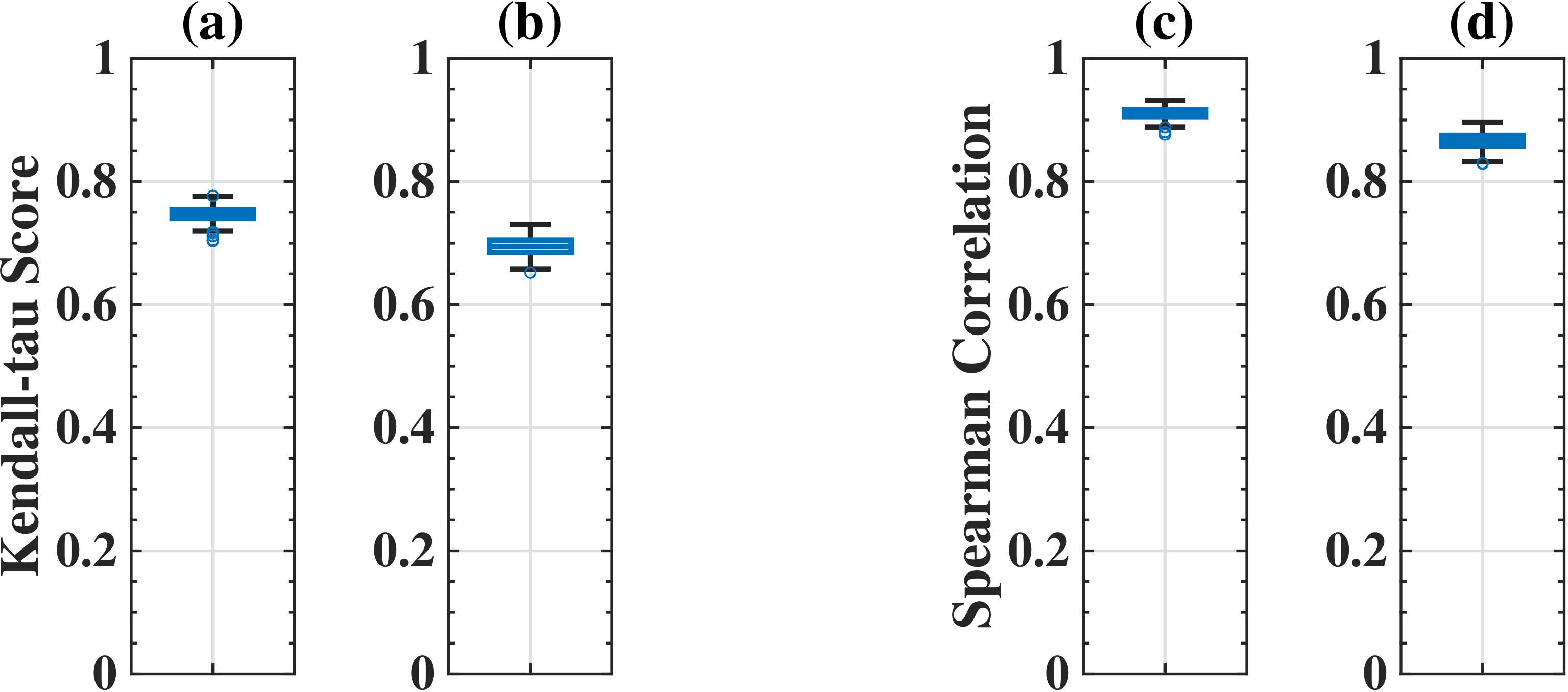}
        \caption{Training and testing ranking score distribution for Minnesota State Transportation network (Case-I). [The Box and Whisker plot shows the median, the lower and upper quartiles, any outliers (calculated using the interquartile range), and the minimum and maximum values that are not outliers]--(a) Kendall-tau score on training data, (b) Kendall-tau score on testing data, (c) Spearman correlation coefficient on training data, (d) Spearman correlation coefficient on testing data.}
        \label{minnesota_wt_change2}
    \end{center}
\end{figure}

For this case, the edge weights are randomly altered according to $r \times w_i$ where $w_i$ is the weight of edge $i$ and the values of $r$ are sampled from the continuous uniform distribution $\mathcal{U}[0.8,1.2]$. The number of nodes and edges for the graphs remain unchanged. The graph statistics are shown in Table \ref{Tab_minnesota_v2_stats}. From this graph generation model, we select 1000 training data and 200 testing data and the scores are shown in Table \ref{Tab_minnesota_v2} and Figure \ref{minnesota_wt_change2}. As for the synthetic graphs, high scores and relatively small standard deviation indicates that the model is fairly robust.

\begin{table}[h!]
\centering
\caption{Statistics of Minnesota State transportation network study (Case-I)}
\begin{tabular}{|ccccc|} \hline
node & edge & Avg. Shortest path lengths & Avg. Clustering coeff. & Average degree \\ 
 nos.       &    nos.     & ($\mu \pm \sigma$, max)    & ($\mu \pm \sigma$)          &  of nodes ($\mu \pm \sigma$)           \\ \hline
2640    & 3302    & $146.5 \pm 0.72$, 148.7  & $0.016 \pm 6.9\times10^{-18}$      & $2.501 \pm 8.88\times10^{-16} $    \\ \hline 
\end{tabular}\label{Tab_minnesota_v2_stats}
\end{table}

\begin{table}[h!]
\centering
\caption{Training and testing ranking scores for Minnesota State Transportation network (Case-I)}
\begin{tabular}{|c|cc|} \hline
                     & Training data                   & Testing data                    \\ \hline
Kendall Tau Score    & \multirow{2}{*}{$0.747 \pm 0.010$} & \multirow{2}{*}{$0.694 \pm 0.014$} \\
($\mu \pm \sigma$)        &                                 &                                 \\
Spearman's Rho Score & \multirow{2}{*}{$0.911 \pm 0.008$} & \multirow{2}{*}{$0.865 \pm 0.013$} \\
($\mu \pm \sigma$)        &                                 &                                \\ \hline 
\end{tabular}\label{Tab_minnesota_v2}
\end{table}


\subsubsection{Case II: Simultaneous change in the edge weights and the number of edges}

\begin{figure}[h!]
    \begin{center}
        \includegraphics[width=0.7\textwidth]{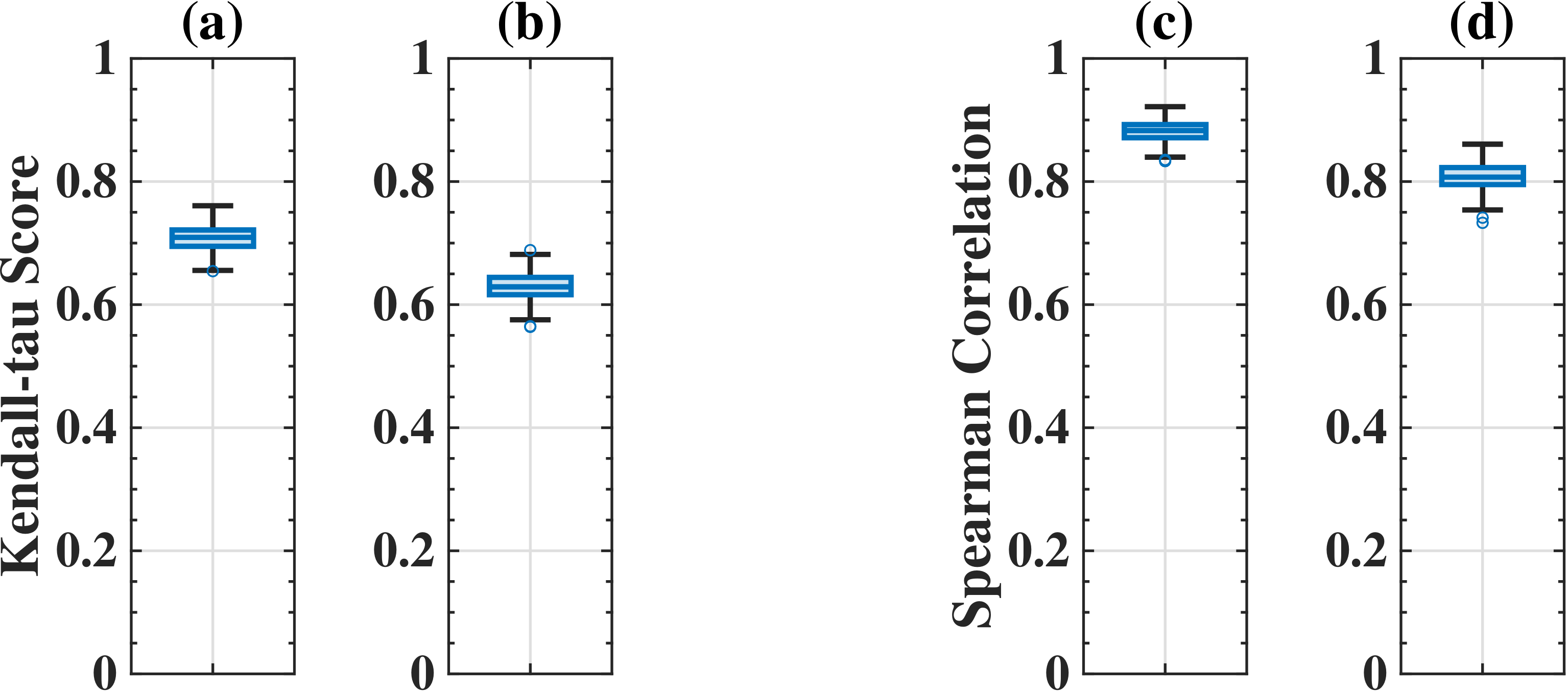}
        \caption{Training and testing ranking score distributions for Minnesota State Transportation network (Case-II). The Box and Whisker plot shows the median, the lower and upper quartiles, any outliers (calculated using the interquartile range), and the minimum and maximum values that are not outliers--(a) Kendall-tau score on training data, (b) Kendall-tau score on testing data, (c) Spearman correlation coefficient on training data, (d) Spearman correlation coefficient on testing data.}
        \label{minnesota_wt_change_node_del2}
    \end{center}
\end{figure}

In this case, the edge weights are as specified in Case-I (Section \ref{Sec_minnesota_case1}) i.e., $r \times w_i$ with $r \in \mathcal{U}[0.8,1.2]$. In addition, the number of edges in the graph are modified. The number of edges is sampled from a discrete uniform distribution $\mathcal{U}\{3269,3302\}$ -- which is based on a maximum of $1\%$ edge deletion from the original network. The case considered, i.e., 1\% edge deletion, reflects an extreme case following a catastrophic event. For example, Shahdani et al. \cite{shahdani2022assessing} found that a major flood event with 20 year return period resulted in 0.6\%-0.7\% non-operating streets in a transportation network. This simulation also includes the addition of new roads. The graph statistics are shown in Table \ref{Tab_minnesota_v6_stats}. After training with 1000 training data, the score is evaluated on 200 test data and training data and shown in Table \ref{Tab_minnesota_v6} and Figure \ref{minnesota_wt_change_node_del2}. Similar to Case-I, the standard deviation of the ranking score is relatively small, indicating the proposed method's robustness.

\begin{table}[h!]
\centering
\caption{Statistics of Minnesota transportation network study (Case-II)}
\begin{tabular}{|ccccc|} \hline
node & edge & Avg. Shortest path lengths & Avg. Clustering coeff. & Average degree \\ 
 nos.       &    nos.     & ($\mu \pm \sigma$, max)    & ($\mu \pm \sigma$)          &  of nodes ($\mu \pm \sigma$)           \\ \hline
2640    & 3269-3302    & $147.6 \pm 1.08$, 151.4  & $0.016 \pm 3.2\times10^{-4}$      & $2.49 \pm 7.2\times10^{-3} $    \\ \hline 
\end{tabular}\label{Tab_minnesota_v6_stats}
\end{table}

\begin{table}[h!]
\centering
\caption{Training and testing ranking scores for Minnesota transportation network (Case-II)}
\begin{tabular}{|c|cc|} \hline
                     & Training data                   & Testing data                    \\ \hline
Kendall Tau Score    & \multirow{2}{*}{$0.708 \pm 0.019$} & \multirow{2}{*}{$0.63 \pm 0.023$}  \\
($\mu \pm \sigma$)        &                                 &                                 \\
Spearman's Rho Score & \multirow{2}{*}{$0.881 \pm 0.015$} & \multirow{2}{*}{$0.808 \pm 0.022$} \\
($\mu \pm \sigma$)        &                                 &                                \\ \hline
\end{tabular}\label{Tab_minnesota_v6}
\end{table}


\section{Conclusions}\label{Sec_Conclusion}

Estimating the importance of the edges in the graph is crucial in many applications, such as in transportation, power, and water distribution networks. The shortest path-based conventional approach can solve the problem at the cost of substantial computational overhead. We show that using Graph Neural Networks (GNNs) to approximate the importance of edges can overcome this issue at the expense of only a moderate reduction in accuracy. Such reductions may be acceptable in emergencies where latency is a primary concern. The computational time of the proposed method is linear order, which beats the quadratic order complexity of the conventional technique. The results of the GNN model in terms of the edge ranking are well correlated to the EBC values obtained using the traditional approach. This approximate estimation of edge ranking is faster for large graph networks, say with node numbers of more than 2000, which is the main advantage. The proposed GNN model is scalable-- the number of edges in the graph can vary and the model can be applied to graphs of varying sizes. It also generalizes, as the validation is performed on data unseen by the model during the training phase. The proposed framework is validated on three types of synthetic graphs: Erd{\H{o}}s-R{\'e}nyi-I, Erd{\H{o}}s-R{\'e}nyi-II, and Watts-Strogatz model. The mean Spearman's correlation coefficient for all those cases on the test dataset is approximately 0.92, which denotes the efficacy of the proposed method. We also studied the proposed framework's performance on the Minnesota transportation network. For this, we considered two cases - (a) change in the edge weights and (b) change in edge weight along with edge deletion/addition. For these cases, the Spearman correlation coefficients are $0.865$ and $0.808$, respectively. Results show that the proposed method has excellent potential for finding important graph components in the field of network systems. From the wide range of graphs used here to validate the approach, we can conclude that there is significant promise for using this approach towards maintenance, recovery time, and resilience estimation for a wide range of networked infrastructure systems.

\section*{CRediT authorship contribution statement}

\textbf{Debasish Jana}: Conceptualization, Methodology, Software, Validation,  Formal analysis, Investigation, Resources, Data curation, Writing – original draft, Visualization.\\
\indent \textbf{Sven Malama}: Conceptualization, Methodology, Software, Validation,  Formal analysis, Investigation, Resources, Data curation, Writing – review \& editing, Visualization.\\
\indent \textbf{Sriram Narasimhan}: Conceptualization, Methodology, Analysis, Investigation, Resources, Writing – review \& editing, Supervision, Project Administration, Funding acquisition.\\
\indent \textbf{Ertugrul Taciroglu}: Conceptualization, Investigation, Resources, Writing – review \& editing, Supervision, Project Administration, Funding acquisition. 

\section*{Declaration of Competing Interest}
The authors declare that there is no conflict of interest regarding financial and personal relationships that could influence the work reported in the paper.

\section*{Funding Information}
We gratefully acknowledge funding support provided by the City of Los Angeles’s Bureau of Engineering (BOE).


\appendix

\section{Evaluation Metric}\label{EvalMets}

\subsection{Kendall tau rank correlation coefficient}

Kendall tau rank correlation coefficient \cite{kendall1938new} is a popular metric for the ranking measure. Let's consider $(x_1, y_1), \cdots, (x_n, y_n)$ are the set of observations of the joint random variables $X$ and $Y$. Any pair of observations $(x_i, y_i)$ and $(x_j, y_j)$, where $i<j$, are said to be concordant if the ranks of the both elements in both pairs agree. If not, they are said to be discordant. For two given lists with $n$ items each, if the number of concordant and discordant pairs are $N_c$ and $N_d$ respectively, then Kendall's ranking coefficient is calculated as,


\begin{equation}
    \tau = \frac{\textrm{number of concordant pairs} - \textrm{number of discordant pairs}}{\textrm{number of ways to choose two items from $n$ items}} = \frac{N_c - N_d}{\frac{n(n-1)}{2}}.
\end{equation}

Normalizing with the number of pair combinations results in the range of the coefficient $\tau$ between $-1 \leq \tau \leq 1$. The value of $\tau$ is 1, -1, 0 if all the pairs are concordant, discordant and uncorrelated, respectively. 


\subsection{Spearman's rank correlation coefficient}

Spearman's rank correlation coefficient \cite{myers2004spearman} is the covariance of the two rank variables divided by the product of their standard deviation. With $n$ number of observations, the $n$ raw scores of variables $x_i$ and $y_i$ are transformed to the ranks $R(x_i)$ and $R(y_i)$ for the joint random variables $X$ and $Y$. Then the Spearman's rank correlation coefficient $\rho$ is expressed as follows:

\begin{equation}
    \rho_s = \frac{\textrm{cov}(R(X), R(Y))}{\sigma_{R(X)}\sigma_{R(Y)}}
\end{equation}

\noindent where, $\textrm{cov}(R(X), R(Y))$ is the covariance of the rank variables; $\sigma_{R(X)}$ and $\sigma_{R(Y)}$ are the standard deviations of the rank variables. If all the $n$ ranks are distinct integers, the spearman's rank correlation can be simplified as:

\begin{equation}
    \rho_s = 1-\frac{6 \sum d_i^2}{n(n^2-1)}
\end{equation}

\noindent where, $d_i = R(x_i)-R(y_i)$ is the difference between the two ranks of each observation, and $n$ is the number of observations. The range of spearman's coefficient $\rho_s$ is $-1 \leq \rho_s \leq 1$. Like the Kendall tau, the spearman correlation, $\rho_s = 1, -1$, and $0$ denote perfectly positive, perfectly negative, and no correlation, respectively.

While the Kendall tau and Spearman's rank correlation coefficients lead to similar results, in practice, Spearman's $\rho_s$ is more popular as a ranking measure. However, Spearman's $\rho_s$ is more sensitive to errors and discrepancies in the data. On the other hand, if the data is Gaussian distributed, Kendall tau has less gross error sensitivity and less asymptotic variance compared to Spearman's $\rho_s$. Therefore, error metrics for all synthetic and experimental simulations are reported in this paper.

\section{Ablation Study}\label{Sec_Ablation}
We performed a suite of experiments on synthetic graphs to study the effect of hyper-parameters on the GNN model's performance. The main hyper-parameters of the model are the number of GNN layers and the number of embedding dimensions. Therefore, we vary these hyper-parameters and observe the performance of the model. We use Erd{\H{o}}s - R{\'e}nyi variant-I (GNP random) graphs for this study.

\subsection{Varying number of layers}\label{Sec_varying_layers}

\begin{figure}[h!]
    \begin{center}
        \includegraphics[width=0.5\textwidth]{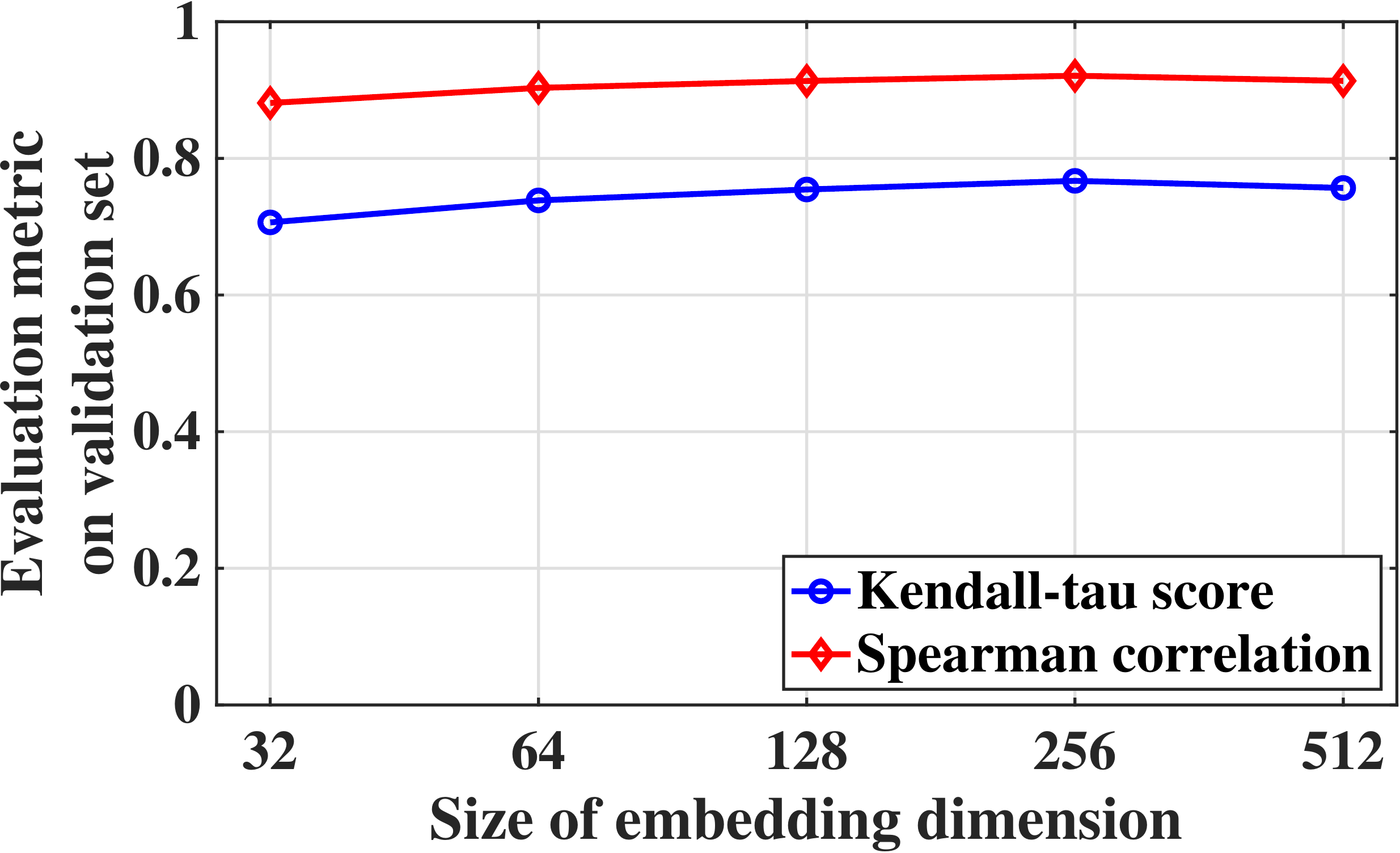}
        \caption{Evaluation metric for different number of GNN layers}
        \label{ablation_layers2}
    \end{center}
\end{figure}

The number of GNN layers in the model influences the amount of information any given edge can accumulate from its neighboring edges. For an increasing number of GNN layers, the edges have access to information from multi-hop adjacent edges. In this study, we vary the number of GNN layers from 1 to 5, keeping the embedding dimension fixed (256). Additionally, we present the model performance in Figure \ref{ablation_layers2}. Both evaluation metrics, i.e., Kendall tau and Spearman's correlation coefficient, show that models with small numbers of GNN layers perform poorly, as the feature aggregation reach for each edge is limited. Increasing the number of GNN layers yields better ranking performance. Therefore, we fix the number of GNN layers as 5 for all the numerical and experimental studies; increasing this number further comes at the cost of higher training time with only a marginal improvement in accuracy.


\subsection{Varying embedding dimensions}

\begin{figure}[h!]
    \begin{center}
        \includegraphics[width=0.5\textwidth]{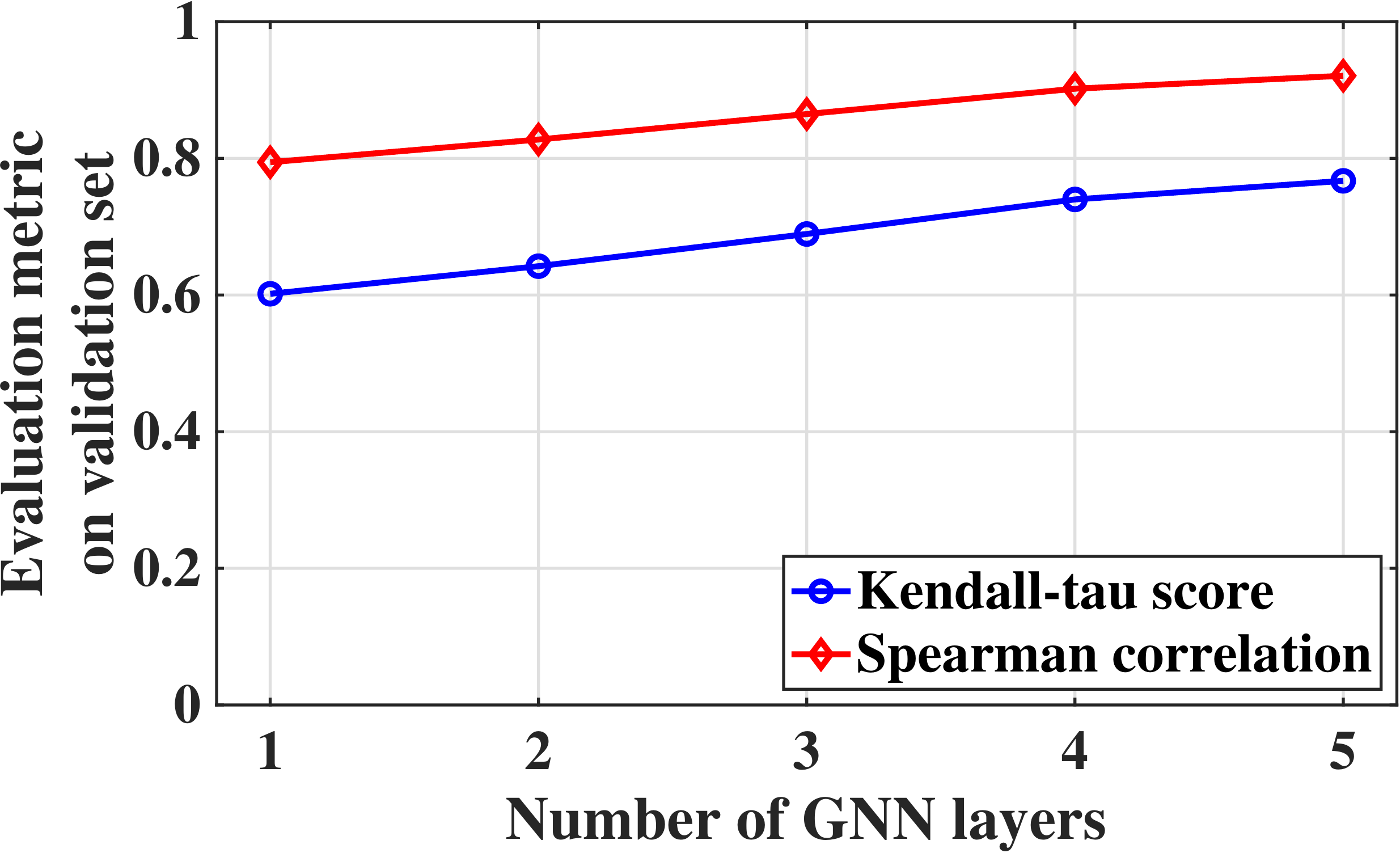}
        \caption{Evaluation metric for different size of embedding dimension}
        \label{ablation_hidden2}
    \end{center}
\end{figure}

For any neural network (shallow or deep), the embedding dimension (number of neurons in the hidden layers) represents the number of learnable modal parameters. Under-parameterized models (low embedding dimension) cannot approximate complex functions, whereas over-parameterized models (high embedding dimension) generalize poorly. In this experiment, we change the embedding dimension to 32, 64, 128, 256, and 512, with five layers (obtained from the \ref{Sec_varying_layers}). We evaluate the performance for all these trials and present it in Figure. \ref{ablation_hidden2}. Results show that an embedding dimension of 256 is optimal -- performance is lower for both the lower and higher embedding dimensions.


\bibliographystyle{elsarticle-num}
\bibliography{references}  






\end{document}